\documentclass[%
  reprint,
  superscriptaddress,
  longbibliography,
  amsmath,amssymb,
  aps,
  prl,
  floatfix,
  nobalancelastpage,
]{revtex4-2}

\usepackage{amsfonts,amssymb,amsmath}
\usepackage{graphicx}
\usepackage{dcolumn}
\usepackage{bm}
\usepackage[mathlines]{lineno}
\usepackage{seqsplit} 
\usepackage{physics}
\usepackage{bbm}
\usepackage{xcolor}
\usepackage{soul}
\usepackage{comment}
\usepackage{amsthm}
\usepackage{float}
\usepackage{grffile}
\usepackage{booktabs}
\theoremstyle{plain}

\theoremstyle{remark}

\theoremstyle{plain}

\theoremstyle{definition}

\theoremstyle{remark}

\usepackage{orcidlink}
\usepackage{microtype}
\usepackage{lipsum}
\usepackage{hyperref}
\graphicspath{{./Figures/}}
\usepackage[normalem]{ulem}
\usepackage{mathrsfs}

\newcommand{\beq}{
\begin{equation}}
  \newcommand{\eeq}{
\end{equation}}

\newcommand{\dg}{^\dagger}

\newcommand{\smallfrac}[2]{\mbox{$\frac{#1}{#2}$}}

\newcommand{\half}{\smallfrac{1}{2}}
\renewcommand{\bra}[1]{\langle{#1}|}
\renewcommand{\ket}[1]{|{#1}\rangle}

\renewcommand{\op}[2]{\ket{#1}\bra{#2}}

\newcommand{\sq}[1]{\left[ {#1} \right]}

\renewcommand{\Tr}{\text{Tr}}

\newcommand{\s}[1]{\hat{\sigma}_{#1}}

\newcommand{\ex}[1]{\langle{#1}\rangle}
\renewcommand{\dd}{{\rm d}}

\newcommand{\ie}{{\em i.e.}}

\newcommand{\incoh}{_{\cal I}}
\newcommand{\coh}{_{\cal C}}

\renewcommand{\sq}{_{\cal S}}
\newcommand{\mx}{_{\mathcal{T}h}}
\renewcommand{\ng}{_{\cal NG}}

\newcommand{\figref}[1]{Fig.~\ref{#1}}
\renewcommand{\eqref}[1]{Eq.~(\ref{#1})}

\definecolor{nblue}{rgb}{0.06,0.3,0.73}
\definecolor{nblack}{rgb}{0,0,0}
\definecolor{nred}{rgb}{0.9,0.1,0.1}
\definecolor{nmagenta}{rgb}{0.7,0.0,0.3}
\definecolor{applegreen}{rgb}{0.55, 0.71, 0.0}
\definecolor{nteal}{rgb}{0, 0.5, 0.5}

\begin{document}

\title{Optimizing Wigner Negativity in Scattering Processes Using Energetic Cost Functions}
\author{Kian Hwee Lim\orcidlink{0000-0003-2154-4288}}
\email{kianhwee.lim95\@gmail.com}
\affiliation{MajuLab, CNRS-UCA-SU-NUS-NTU International Joint Research Laboratory}
\affiliation{Centre for Quantum Technologies, National University of Singapore, 117543 Singapore, Singapore}
\affiliation{CNRS@CREATE LTD, 1 Create Way, \#08-01 CREATE Tower, Singapore 138602}
\author{Kiarn T. Laverick\orcidlink{0000-0002-3688-1159}}
\affiliation{MajuLab, CNRS-UCA-SU-NUS-NTU International Joint Research Laboratory}
\affiliation{Centre for Quantum Technologies, National University of Singapore, 117543 Singapore, Singapore}
\author{Sahil S. Jafar \orcidlink{0009-0007-2011-3669}}
\affiliation{MajuLab, CNRS-UCA-SU-NUS-NTU International Joint Research Laboratory}
\affiliation{Centre for Quantum Technologies, National University of Singapore, 117543 Singapore, Singapore}
\affiliation{School of Physical and Mathematical Sciences, Nanyang Technological University, 21 Nanyang Link, Singapore 637371, Singapore}
\author{Samyak P. Prasad\orcidlink{0009-0001-3672-6001}}
\affiliation{MajuLab, CNRS-UCA-SU-NUS-NTU International Joint Research Laboratory}
\affiliation{Centre for Quantum Technologies, National University of Singapore, 117543 Singapore, Singapore}
\author{Maria Maffei\orcidlink{0000-0001-5183-4716}}
\affiliation{Universit\'e de Lorraine, CNRS, LPCT, F-54000 Nancy, France}
\author{Alexia Auff\`eves\orcidlink{0000-0003-4682-5684}}
\email{alexia.auffeves@cnrs.fr}
\affiliation{MajuLab, CNRS-UCA-SU-NUS-NTU International Joint Research Laboratory}
\affiliation{Centre for Quantum Technologies, National University of Singapore, 117543 Singapore, Singapore}

\date{\today}

\begin{abstract}
  Wigner negativity is a key resource for quantum technologies but is difficult to optimize in multimode scattering systems. We study the scattering of coherent pulses by a two-level emitter coupled to a one-dimensional waveguide and introduce energetic cost functions that enable the optimization of Wigner negativity without reconstructing the full Wigner function. By decomposing the scattered energy into coherent, thermal, squeezing, and non-Gaussian contributions, we identify an energetic witness that strongly correlates with the achievable negativity across all driving regimes. This approach singles out optimal output temporal modes and uncovers operating points generating appreciable Wigner negativity with sub-photon input energies. We further identify a maximal energy-efficiency regime at spectral mode matching, where the emitter effectively implements a vacuum-selective $\pi$ phase shift, realizing a giant optical nonlinearity. These results establish energetic optimization as a practical route to engineering Wigner-negative photonic states in waveguide quantum electrodynamics and related bosonic scattering platforms.
\end{abstract}
\maketitle

{\em Introduction.}---Wigner negativities (WNs) signal the non-Gaussian character of bosonic states \cite{negativity_measure,albarelli2018resource}. Beyond their fundamental interest~\cite{liu2022experimental, walschaers2023quantumsteering, booth2022contextuality}, they constitute key resources for quantum metrology and sensing \cite{adesso2009optimal,genoni2009enhancement,tan2019nonclassical,rahman2025genuine}, as well as for continuous-variable quantum computing~\cite{kenfack2004negativity,mari2012positive,veitch2013efficient,lvovsky2020production}. Generating WNs from initially Gaussian states requires nonlinear processes, which can arise, for instance, from quantum measurements \cite{genoni2010non,wang2025limitations} or from scattering off a quantum emitter—the simplest example being a two-level atom (TLA) coupled to a one-dimensional (1D) reservoir of bosonic modes \cite{he2013demand}.
Scattering processes involving such ``1D atoms" are particularly appealing because they can be treated as unitary and intrinsically lossless processes \cite{maffei2022closed-system,prasad2026thermodynamics}. As a result, they are deterministic and energy-preserving. These schemes can be realized in state-of-the-art platforms in quantum photonics~\cite{scarpelli2019betafactor, siampour2017nanofabrication, huet2026industryready} and circuit quantum electrodynamics (cQED)~\cite{blais2021circuit}, where they already enable the implementation of quantum interfaces~\cite{javadi2018spin, sipahigil2016an}, quantum gates \cite{joshi2021quantum,krastanov2022controlled,pettersson2026high}, and the generation of resource states~\cite{hu2008deterministic,hoi2012generation, javadi2015single}. While classical $\pi$-pulses based on short and intense input pulses are natural candidates to generate WNs through the spontaneous emission of single photon states, recent studies have focused on drives of finite intensities in the steady-state scattering regime~\cite{quijandria2018steadystate, strandberg2019numerical, kleinbeck2023creation}. They have demonstrated the emergence of WNs in the resonance fluorescence of a 1D atom, with typical negativities on the order of $10\%$ that of a single photon's negativity. This motivates to develop a physical understanding of the mechanisms yielding WNs beyond the $\pi$ pulse regime, as well as strategies to optimize them.

\begin{figure}[h]
  \centering
  \includegraphics[width=0.48\textwidth, trim={0cm 0cm 0cm 0cm},clip]{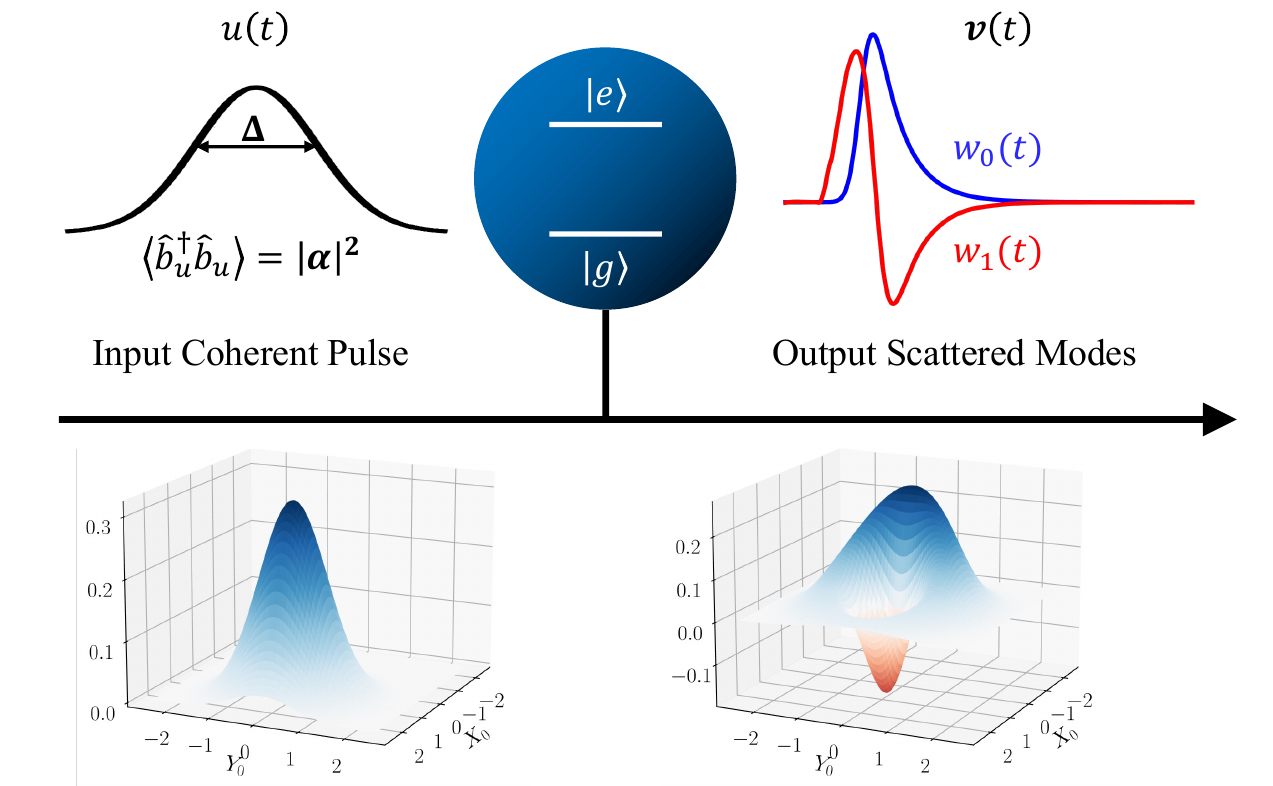}
  \caption[]{
    An input coherent pulse of Gaussian temporal shape $u(t)$, typical width $\Delta$ and
    mean photon number $|\alpha|^2$ scatters off a two-level atom, producing
    multi-modal output light, represented by the
    temporal modes $w_0$ and $w_1$.
    For each $(\alpha, \Delta)$, a carefully designed output filter function
    $v(t)$ selects from the multi-modal output light a temporal mode $v$ with Wigner negativity $\mathcal{N}^v$.
  }
  \label{fig: fig1}
\end{figure}

This is the purpose of this Letter, which focuses on the scattering of a coherent pulse by a 1D atom. Our study encompasses finite energy pulses allowing for incoherent scattering to take place while the TLA is driven. Here optimizations are particularly challenging because the scattered field is inherently multimode: consequently, they must be performed not only over the input pulse, but also over the output modes in which the WN is detected~\cite{brecht2015photon, loudon2000quantum}. In addition, evaluating WNs requires prior computation of the Wigner function over the full phase space of the detected mode, making it impractical as a cost function for direct optimization. To overcome this difficulty, we exploit the energy-preserving nature of scattering processes and instead introduce optimization strategies based on energetic cost functions.

The central idea of our approach is to identify within the energy of the
scattered field the components which do not contribute to WNs. By maximizing the
remnant energetic contribution over all possible output modes, we can single out the
output mode most likely to yield a large WN. We implement this
optimization, first using as a probe the energy of the field fluctuations—or incoherent energy
\cite{laverick2026energetic}—and then refine the analysis by isolating in the
incoherent energy a specific contribution we attribute to the non-Gaussian (NG)
character of the field~\cite{referenceToSahilPaper}. Varying the width and the intensity of the input pulse, we observe a strong correlation
between the maxima of the WN and those of the NG energetic contribution, over all regimes of light-matter coupling. In addition to the expected $\pi$ pulses, our study reveals new operating points yielding sensible WNs, based on input pulses that contains less than one photon on average. We show that the most energy-efficient point is
spectrally mode-matched with the TLA. In this regime, the
TLA selectively induces a $\pi$ phase shift on the vacuum component, i.e. behave as a giant optical non-linear medium~\cite{kojima2004efficiencies,GNL2007}. These results demonstrate that energetic cost functions provide an efficient route to identifying input and output pulses that
maximize WNs, while unveiling key mechanisms yielding the
generation of NG states under energy constraints.
\\

\noindent {\em System and model.---}We consider a TLA coupled to a 1D reservoir of electromagnetic modes labelled by their frequency $\omega$ (See Fig.~\ref{fig: fig1}). The total Hamiltonian reads (in units of $\hbar$) $\hat{H} = \hat{H}_{\rm Q} + \hat{H}_{\rm F}+
\hat{H}_\mathrm{int}$. The TLA Hamiltonian is $\hat{H}_{\rm Q} = \omega_{{q}} \s{}^\dagger \s{}$, with $\s{}$ the TLA lowering operator, the field Hamiltonian is $\hat{H}_{\rm F} = \int
\dd \omega \, \omega \hat{b}^\dagger(\omega)\hat{b}(\omega)$, with $\hat{b}(\omega)$ the field annihilation operator for the mode of frequency $\omega$, verifying $[\hat{b}(\omega),\hat{b}^{\dagger}(\omega')]=\delta(\omega-\omega')$, and the interaction term is $\hat{H}_\mathrm{int} = i\sqrt{\frac{\gamma}{2\pi}} \int \dd\omega \, \left(\hat{b}^\dagger(\omega) \s{} - \s{}^\dagger \hat{b}(\omega)\right)$ where $\gamma$ refers to the TLA's spontaneous emission rate. We introduce the temporal mode operator $\hat{b}(t)=(2\pi)^{-1/2}\int d\omega e^{i(\omega-\omega_q)t}\hat{b}(\omega)$ that verifies $[\hat{b}(t),\hat{b}^{\dagger}(t')]=\delta(t-t')$. The joint evolution of the field state and of the TLA is computed in the interaction picture with respect to $\hat{H}_Q+\hat{H}_F$.

In this work, we consider coherent input pulses $\ket{\alpha_\text{in}} =
e^{\alpha \hat{b}_u^\dagger - \alpha^* \hat{b}_u  }\ket{\text{vac}}$, where
$\alpha$ is a complex number and $\ket{\text{vac}}$ the vacuum of the 1D modes.
We have introduced the input mode $\hat{b}_u = \int \dd t\, u(t)
\hat{b}(t)$, where $u(t) = (\pi \Delta^2)^{-1/4} {\rm
exp}[-(t-t_0)^2/2\Delta^2]$ is a normalized Gaussian function, such that
$\hat{b}_u$ is a bosonic operator.  $(\alpha,\Delta)$ fully parametrize the input
pulse, $|\alpha|^2$ being the mean number of photons in the pulse and $\Delta$
its typical duration.
If $\gamma \Delta \ll |\alpha|^2$ ($\gamma \Delta \gg
|\alpha|^2$) stimulated (spontaneous) emission dominates during the TLA-pulse interaction. The stimulated regime provides a clear temporal separation
between the coherent excitation of the TLA, and the spontaneous release of a field in a single temporal mode. In full generality however, incoherent scattering takes place while the TLA is driven.
In this case a few to several output modes are
populated and possibly entangled \cite{carmichael2013statistical}, while only one mode is detected and characterized. The detected mode is parametrized by the normalized filter function $v(t)$ and the related temporal
mode operator $\hat{b}_v = \int \dd t\, v(t) \hat{b}(t)$
\cite{kiilerich2020quantum}.   Once the input and output parameters are set, it is possible to compute the evolved quantum state of the field inside the detected
mode $v$ using the virtual cavity
approach~\cite{kiilerich2020quantum}, so as its Wigner function $W^v(\beta)$ and WN, ${\cal N}^v = \int \dd^2\beta |W^v(\beta)| - 1$ (See end matter).  \\

\begin{figure}
  \centering
  \includegraphics[width=0.5\textwidth, trim={0cm 0cm 0cm 0cm},clip]{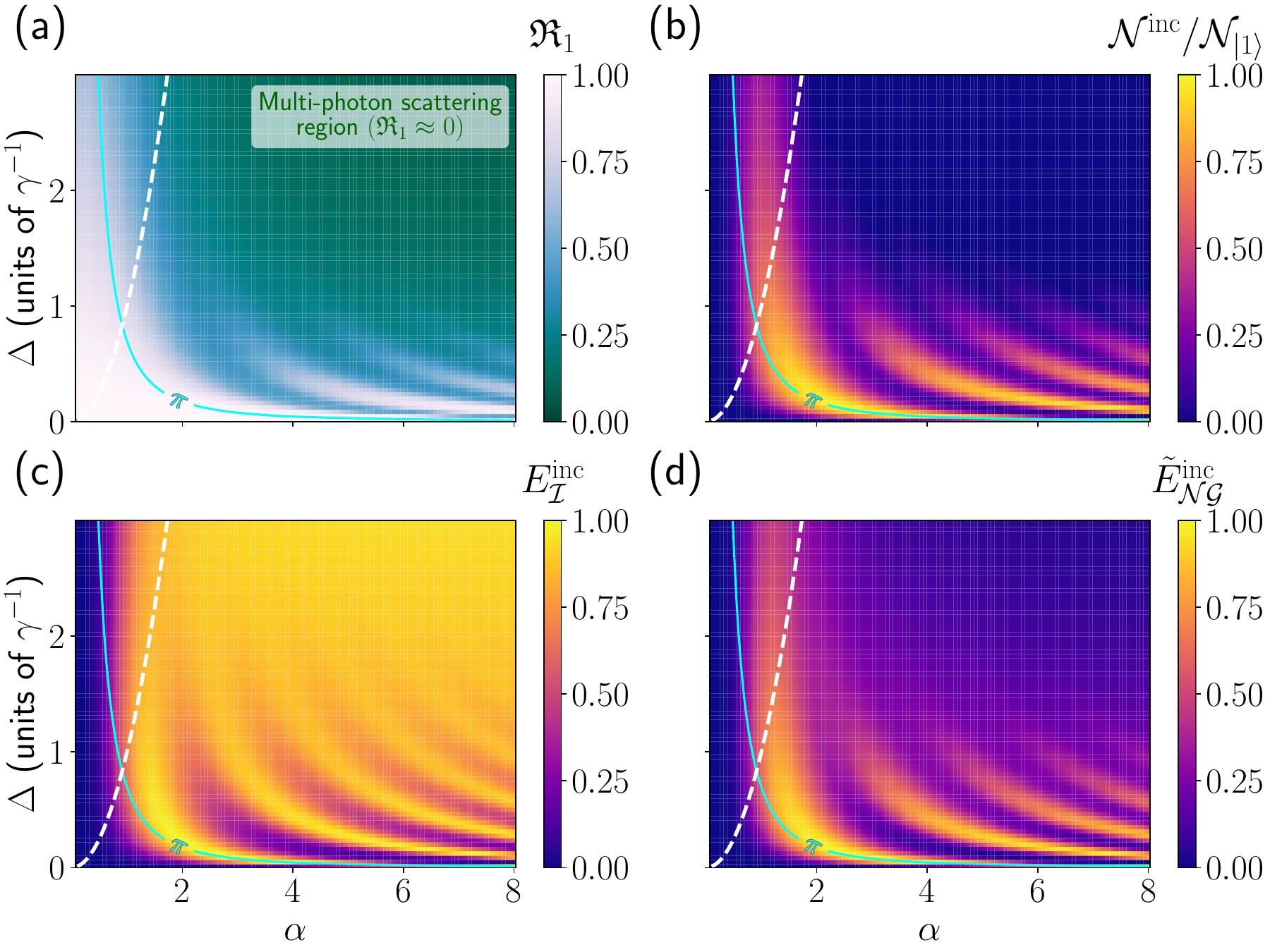}
  \caption[]{ \textbf{(a)}~Rank-one-score of the incoherent modes's basis
    [see~\eqref{eqn: rank-one score defn}]. {\textbf{(b)}}~Wigner negativity $\mathcal{N}^\text{inc}$,
    {\textbf{(c)}}~$E^\text{inc}\incoh$, and
    {\textbf{(c)}}~$\tilde{E}^\text{inc}\ng$ of the field in the most
    incoherent mode $w_\text{inc}$. The cyan contour line corresponds to a $\pi$-pulse (See text). The
  white dashed-line typically separates the spontaneous from the stimulated emission regimes (See text).}
  \label{fig: fig2}
\end{figure}

\noindent {\em Incoherent modes' basis.}---From the scenario above, it is clear that a direct optimization of the detected
WN as a function of the parameters $(\alpha, \Delta, v(t))$ is highly
impractical, calling for alternative strategies. In this section we fix $(\alpha,\Delta)$ and focus on the filter mode $v$. Our approach is based on the simple intuition that only the fluctuations of the mode contributes to WNs. Namely, counting $E_v$ in photon number (see end matter), we split it into two components $E_v = E^v\coh + E^v\incoh$ \cite{referenceToSahilPaper,laverick2026energetic}. The {\em coherent} energy $E^v\coh := |\langle \hat{b}_v \rangle|^2$ quantifies the mean field energy. The {\em incoherent} energy $E^v\incoh := \ex{\delta \hat{b}_v\dg\delta \hat{b}_v}$,
with $\delta \hat{A}:= \hat{A}-\langle \hat{A}\rangle$, quantifies the energy locked in the field fluctuations. $E^v\coh$ does not yield WNs as any change in $E^v\coh$ amounts to a displacement of the field: thus we rather search for filter modes with large $E^v\incoh$. To do so, we define the correlation function of the output field's fluctuations, $ K^{\cal I}(t, t^\prime) := \langle \delta
\hat{b}^\dagger_{\text{out}}(t) \delta \hat{b}_{\text{out}}(t^\prime) \rangle$, such that $E^v\incoh = \iint \dd t
\dd t^\prime \, v^*(t) K^\mathcal{I}(t,t^\prime) v(t^\prime)$ (see end matter).
$\hat{b}_{\text{out}}$ is the output operator defined in the
Heisenberg picture~\cite{gardiner1985input}. $K^\mathcal{I}(t,
t^\prime)$ being positive, semi-definite, and upper bounded by $|\alpha|^{2}$,
the spectral theorem~\cite{conway2007operators} states there exists an
orthonormal basis $\{ w_j(t) \}_{j\geq 0} $ of filter functions such that
\begin{align}
  \label{eqn: Kdelta}
  K^\mathcal{I}(t, t^\prime) & = \sum_{j}^{}\delta n_j w_j(t) w_j^*(t^\prime)\,.
\end{align}
$\delta n_j$ are real numbers referring to the mean number of incoherent photons
in the mode defined by $w_j(t)$, with $\delta n_i \geq \delta n_j$ for $i < j$: $w_0:=w_\text{inc}$ thus captures the most incoherent mode we shall focus on shortly. $\delta n_j \leq 1$, which reflects the TLA's
inability to incoherently scatter more than one photon in any given mode (see
end matter).  The basis of
incoherent modes is characterized by its {\em rank-one score}:
\begin{equation}
  \label{eqn: rank-one score defn}
  {\mathfrak R}_1 := \frac{\sum_{j=0} (\delta n_j)^2}{\left(\sum_{j=0} \delta n_j\right)^2}\,.
\end{equation}
${\mathfrak R}_1 < 1$ quantifies the spreading of the incoherently scattered
field over various modes. ${\mathfrak R}_1 = 1$ if the fluctuations of the scattered field is contained in the single mode $w_0$.  ${\mathfrak R}_1$ is depicted in Fig.~\ref{fig: fig2}(a) as a function of the input parameters $(\alpha,\Delta)$.  The dotted line $\Delta \gamma = |\alpha|^2$ typically separates the spontaneous from the stimulated emission regime. The stimulated region is characterized by oscillations of ${\mathfrak R}_1$, which reflect the coherent oscillations of TLA's population. So called $(2k+1)\pi$-pulses (see S.1 in
SM~\cite{referenceToSupplementalMaterial}) induce the TLA's population inversion, followed by the release of a single photon in a single temporal mode (${\mathfrak R}_1 \sim 1$). Large values of ${\mathfrak R}_1$ are also reached for long, low energy pulses ($\gamma \Delta \gg 1$ and $\alpha \leq 1$): in this overdamped regime, photons are preferentially scattered in the exciting mode, which is the approximately flat temporal mode.  Out of these specific regions,
${\mathfrak R}_1$ quickly decreases because the field is scattered in more than one mode (multi-photon emission).\\

\noindent {\em Most incoherent mode.}---From now on we fix our filter to be on the most incoherent mode $w_{\text{inc}}(t)$. We focus on the evolution of the WN captured in this mode, denoted $\mathcal{N}^{\text{inc}}$, as a function of the input parameters $(\alpha,\Delta)$ (Fig.~\ref{fig: fig2}(b)). WNs are normalized with respect to a single photon's negativity, $\mathcal{N}_{\ket 1} = 4e^{-1/2} - 2$ \cite{kenfack2004negativity}. As it appears on the figure, the temporal mode filter $v(t)$ perfectly captures maximal WNs ($\mathcal{N}^{\text{inc}}/\mathcal{N}_{\ket 1}=1$) along sharp $\pi$ pulses, which release single photons. It also unveils a region of low energy input pulses ($\alpha \sim 1$) giving rise to sensible WNs - the mechanism behind these energy-efficient points will be analyzed below. For the sake of comparison, we provide a similar plot taking $v(t)$ as the most populated mode (see end matter). We observe that $\pi$ pulses are not captured by that approach.

We now explore the potential of the incoherent energy to single out optimal {\it input} parameters. $E^{\text{inc}}\incoh$ is plotted as a function of $(\alpha,\Delta)$ on Fig.~\ref{fig: fig2}(c). Like $\mathcal{N}^{\text{inc}}$, $E^{\text{inc}}\incoh$ is maximized along sharp $\pi$ pulses, where it trivially reaches $1$ for single photon emission. However contrary to $\mathcal{N}^{\text{inc}}$, it quickly saturates in the multi-photon region ($E^{\text{inc}}\incoh \sim 1$). This is because the populated modes can become entangled and squeezed~\cite{carmichael2013statistical}: hence the fluctuations in $w_\text{inc}$ stem from the mixed and squeezed nature of the detected field, rather than from its eventual non-Gaussianity. This invites us to search for other energetic probes.

To do so, we introduce a fine structure to the
incoherent energy in the temporal mode $v$, $E^v\incoh = E^v\sq +  E^v\mx +
\tilde{E}^v\ng $ (see end matter for exact expressions and Ref.~\cite{referenceToSahilPaper} for complete scientific justification in the case of single mode cavities). $E^v\sq$
($E^v\mx$) captures an energetic contribution due to the squeezed (mixed) nature of the field state. $\tilde{E}^v\ng$ is a remnant contribution to $E^v\incoh$. It was shown to be a measure of non-Gaussianity if $v$ hosts a pure state, and a witness if not (See Ref.~\cite{referenceToSahilPaper}). It is plotted on \figref{fig: fig2}(d) (see \figref{fig: wincEsqAndnth} in the end matter for
$E^\text{inc}\sq$, $E^\text{inc}\mx$). We observe a remarkable correlation between the maxima of $\tilde{E}^\text{inc}\ng$ and $\mathcal{N}^{\text{inc}}$ in the whole input parameters's space. This shows that $\tilde{E}^v\ng$ is a good probe to maximize WNs as a function of the input parameters.  \\

\begin{figure}
  \centering
  \includegraphics[width=0.48\textwidth, trim={0cm 0cm 0cm 0cm},clip]{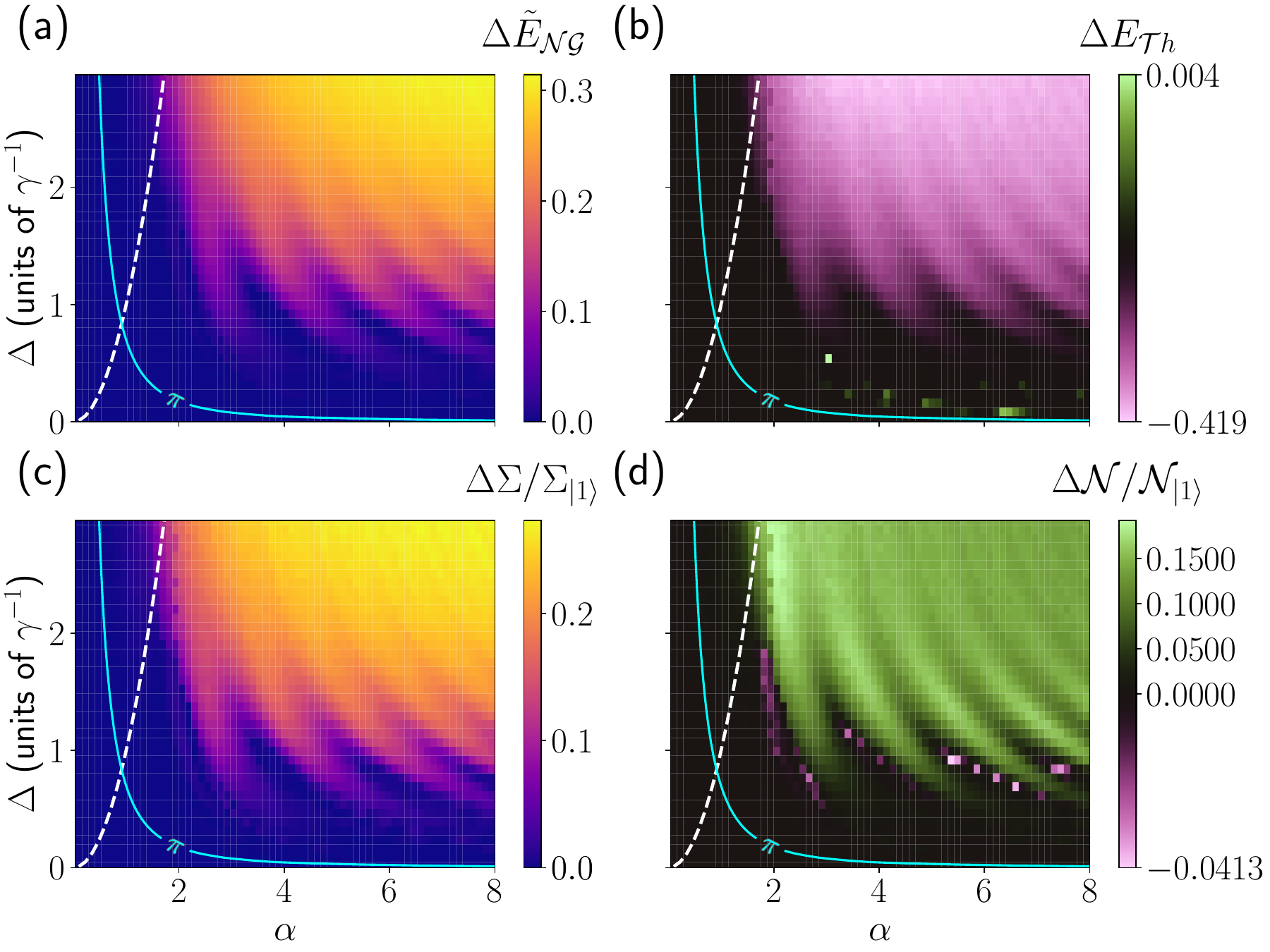}
  \caption[]{ \textbf{(a)}~$\Delta
    \tilde{E}\ng$,  \textbf{(b)}~$\Delta E\mx$,
    \textbf{(c)}~$\Delta \Sigma /\Sigma_{\ket{1}}$, \textbf{(d)}~$\Delta \mathcal{N}/\mathcal{N}_{\ket{1}}$, as a function of input parameters $(\alpha,\Delta)$ (See text). All the plots are done on a coarser grid as compared to
    \figref{fig: fig2}.
    The white and cyan contours are as in Fig.~\ref{fig: fig2}.
  }
  \label{fig: fig3}
\end{figure}

\noindent {\em Most non-Gaussian mode.---}We finally explore the potential of $\tilde{E}^v\ng$ to single out a better filter mode than $w_\text{inc}$. Using the functions $w_j(t)$ defined in \eqref{eqn: Kdelta} as a test basis, we construct a filter function $w(t) = \sum_j c_j w_j(t)$, where $\sum_j |c_j|^2 = 1$.  We then
optimize over $\{c_j\}_j$ to find the filter mode with the largest NG energetic
contribution, which we denote as $ w_{\text{ng}}(t)$ (see S.2 in
SM~\cite{referenceToSupplementalMaterial} for details). $w_{\text{ng}}$ and its related operator $\hat{b}_{\text{ng}}$
define the ``most non-Gaussian mode'' with associated energetic quantities
$\tilde{E}^\mathrm{ng}\ng, E^\mathrm{ng}\mx,
E^\mathrm{ng}\sq$ and negativity $\mathcal{N}^\text{ng}$. To quantify the advantage of using $w_{\text{ng}}(t)$ over
$w_{\text{inc}}(t)$, we define $\Delta X := X^\text{ng} - X^\text{inc}$ with $X = \tilde{E}\ng, \mathcal{N}, E\mx, \Sigma$. $\Sigma$ is the relative entropy of non-Gaussianity~\cite{paulina2013relative,genoni2008quantifying}, a textbook
measure of the non-Gaussianity which we shall normalize by $\Sigma_{\ket{1}} = \log(4)$. These quantities are shown in \figref{fig: fig3} as a function of $(\alpha,\Delta)$. As expected, $\tilde{E}^v\ng$ and $\Sigma^v\ng$ increase while switching from the most incoherent to the most NG mode. Both $\Sigma$ and $\tilde{E}\ng$ visibly increase in the multi-photon region, which also corresponds to a drastic reduction of the filter mode mixedness, $\Delta E\mx < 0$. Hence in this region, the improvement in the capture of NG states stems from a purification of the filter mode. Conversely, no improvement is obtained in the regime of short pulses ($\gamma \Delta \leq 1$) or low energy ($|\alpha|^2 \leq 1$). In these regions indeed, the most incoherent mode and the most NG mode largely overlap $ w_{\text{ng}}\sim w_{\text{inc}}$ (see end matter), signaling that incoherence mostly stems from non-Gaussianity.

The WN follows the same global trend (see \figref{fig: fig3}d): it increases in the region of multi-photon emission, bringing it to values as high as $15\%$ of a single photon negativity - such typical values were already noticed in the steady-state scattering regime~\cite{quijandria2018steadystate, strandberg2019numerical, kleinbeck2023creation}. Our energetic strategy provides a physical ground to these results. Conversely as expected, the region where $ w_{\text{ng}}\sim w_{\text{inc}}$ brings little change. We even notice a slight decrease of negativity (a few percents) for a few specific points in the vicinity of coherent population inversions. This reflects that mixed NG states are not necessarily Wigner-negative \cite{Walschaers2021PRXQ}. For all practical purposes however, these results confirm the interest of using $\tilde{E}\ng$ as a cost function to optimize WN with respect all optimizing parameters.
\\

\begin{figure}
  \centering
  \includegraphics[width=0.48\textwidth, trim={0cm 0cm 0cm 0cm},clip]{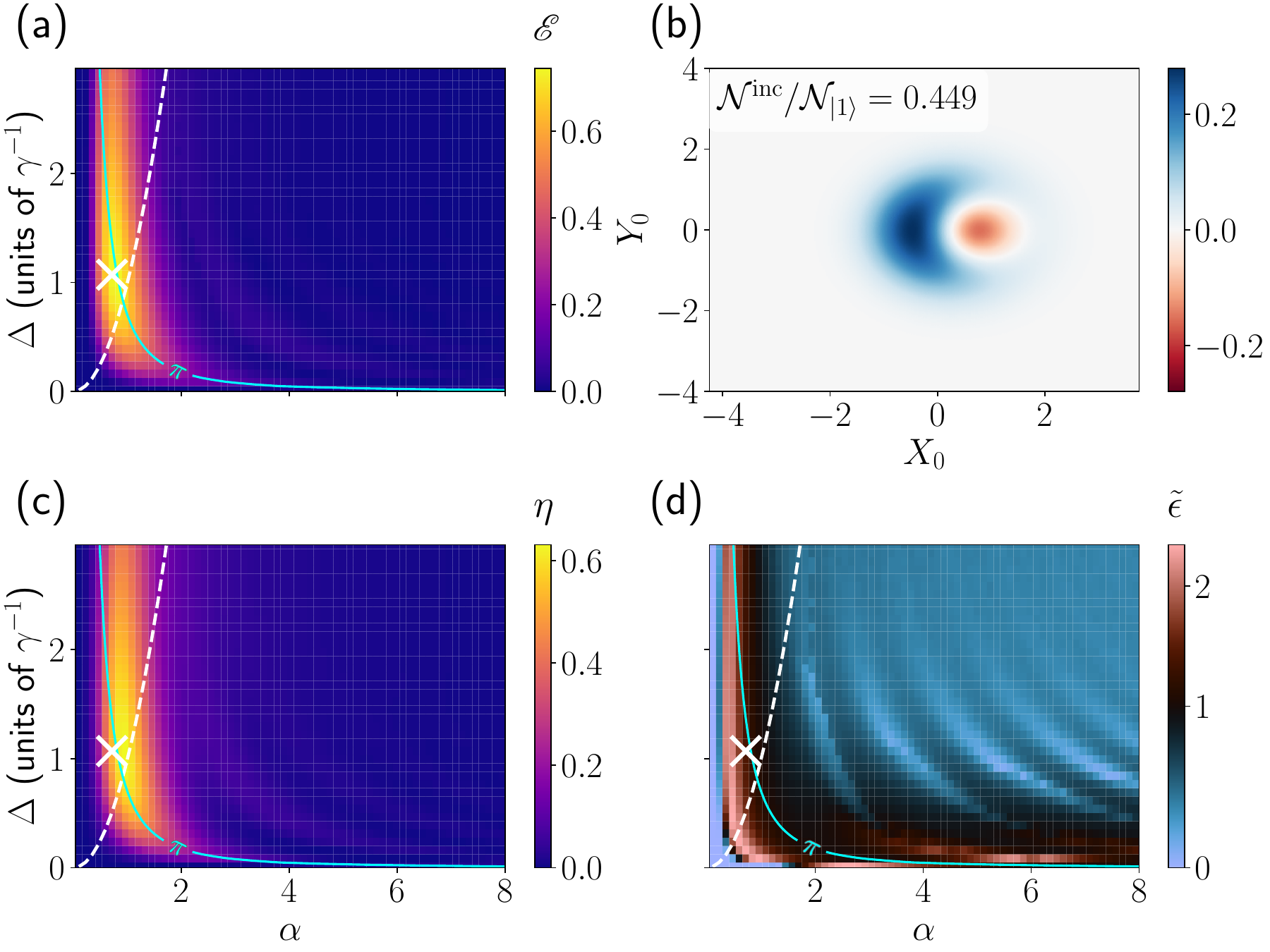}
  \caption[]{\textbf{(a)} $\mathscr{E}^\text{ng}$, \textbf{(c)} $\eta$ and \textbf{(d)} $\tilde{\epsilon}$ as functions of $(\alpha,\Delta)$.  The white cross marks the most energy-efficient point $\mathscr{E}^\text{ng}$ [see
    \eqref{eqn: cascaded efficiency}]. The white and cyan contours
    are as in Fig.~\ref{fig: fig2}.  \textbf{(b)} Wigner function of the output field in temporal mode $w_{\rm ng}$ at the max efficiency
    point.
  }
  \label{fig: fig4}
\end{figure}

\noindent {\em Energy-efficiency of WN generation.---}Above we confirmed $\pi-$pulses as physical processes yielding the
maximal achievable WNs through single photon emission. They also reveal operating points of lower energy ($|\alpha|^2 \sim 1$) where sensible WNs can be generated. This invites to explore the energy-efficiency
of WN generation in the most non-Gaussian mode, defined as
\begin{equation}
  \label{eqn: cascaded efficiency}
  \mathscr{E}^\text{ng} := \frac{\mathcal{N}^\mathrm{ng}/\mathcal{N}_{\ket{1}}}{|\alpha|^2}.
\end{equation}
$\mathscr{E}^\text{ng}$ is a composite figure of merit capturing how much negativity (counted in single photon negativity) is generated per average input photon. Thus it is not necessarily bounded by $1$. $\mathscr{E}^\text{ng}$ is plotted as a function of the input parameters on \figref{fig: fig4}(a), delimitating a region of the phase-space which does not display single photon emission. This is expected since $\pi$ pulses rely on high inputs of energy. We have fully characterized the most efficient point, which reaches $\mathscr{E}^\text{ng} = 0.751$ for $\gamma \Delta =1.070$ (spectral mode-matching) and $\alpha = 0.774$ (onset of the TLA's saturation). The total output field state can be approximated by $\sim |\psi \rangle_{w_{\rm ng}} \otimes |\alpha^\prime \rangle_{\perp}\,$  where $|\alpha^\prime\rangle_{\perp}$ is the coherent field radiated by the TLA in the orthogonal mode $w_\perp$ (see SM~\cite{referenceToSupplementalMaterial}, S.3, for more details).  The Wigner function of $\ket{\psi}_{w_\mathrm{ng}}$ is plotted in \figref{fig: fig4}(b).  Defining $\ket{\beta}$ as a coherent state with $|\beta|^2 = \langle\hat{b}^{\dagger}_{w_{\rm ng}}\hat{b}_{w_{\rm ng}}\rangle$ photons, one finds a high fidelity of $0.986$ between $\ket{\psi}_{w_\mathrm{ng}}$ and
$e^{-i \pi \op{0}{0}} \ket{\beta}$. (See S.4 in
  SM~\cite{referenceToSupplementalMaterial} for a qualitative comparison of their
Wigner functions). Remarkably, this means that the TLA behaves as a giant optical non-linearity, a functionality that was already exploited in \cite{GNL2007,kojima2004efficiencies,wang2011efficient,stobinska2009perfect} to implement photon-photon gates.

To deepen our understanding, we analyze the energy efficiency of WN as a product of two terms, $ \mathscr{E}^\text{ng} = \eta \tilde{\epsilon}$, with $\eta = \tilde{E}^\text{ng}\ng / |\alpha|^2$ and $\tilde{\epsilon} = (\mathcal{N}^\mathrm{ng}/\mathcal{N}_{\ket{1}})/ \tilde{E}^\text{ng}\ng$. $\eta \leq 1$ captures how efficiently the input pulse is funneled into the most non-Gaussian mode. $\tilde{\epsilon}$ is a conversion ratio quantifying the amount of negativity produced per average non-Gaussian photon. Both quantities are plotted on \figref{fig: fig4}(c-d). $\eta$ follows a similar pattern as $\mathscr{E}^\text{ng}$, showing that WN generation efficiency is mostly governed by NG state generation efficiency. Remarkably, the study reveals a regime of parameters where $\tilde{\epsilon}$ can be as high as $2$, meaning that one average NG photon can yield significantly more WN than a single photon. This is the case, e.g. for the optimal point studied above, where $\tilde{\epsilon} = 1.293$. The study of this enhancement will be the subject of future work.\\

{\em Conclusion.}---We have introduced an energetic framework for optimizing Wigner negativity generated in scattering processes and demonstrated its power on the paradigmatic problem of a coherent pulse interacting with a waveguide-coupled two-level atom. By identifying and maximizing the non-Gaussian contribution to the scattered energy, we circumvent the need for repeated Wigner-function reconstructions while efficiently selecting both the optimal detection mode and the optimal driving conditions. Beyond recovering the expected single-photon generation regime associated with $\pi$ pulses, our approach reveals a distinct low-energy operating regime where appreciable Wigner negativity is produced with sub-photon input energies. The optimum energy-efficiency point corresponds to spectral mode matching, where the emitter effectively implements a vacuum-selective $\pi$ phase shift, highlighting the connection between Wigner negativity generation and giant optical nonlinearities. More broadly, these results establish energetic cost functions as practical tools for navigating the large parameter spaces of multimode quantum scattering problems. Extending this approach to multi-emitter architectures, strongly correlated photonic systems, and resource-oriented optimization of other non-classical features may provide new routes toward the efficient engineering of quantum states for continuous-variable processing, quantum networks, and quantum-enhanced technologies.
\\

{\em Acknowledgments}--- The authors warmly thank Hanna Le Jeannic and Thomas Copie for fruitful exchanges. This project is supported by the National Research Foundation, Singapore, through the National Quantum Office, hosted in A*STAR, under its Centre for Quantum Technologies Funding Initiative (S24Q2d0009); by the Plan France 2030 through the projects NISQ2LSQ (Grant ANR-22-PETQ-0006) and OQuLus (Grant ANR-22-PETQ-0013); by the CNRS@ CREATE internal grant ``NGAP'' (NRF2023-ITC004-001); and by OECQ through BPI France. This research is supported by the National Research Foundation, Prime Minister's Office, Singapore under its Campus for Research Excellence and Technological Enterprise (CREATE) programme.

\bibliography{references.bib}

\clearpage

\onecolumngrid

\begin{center}
  \textbf{\large End Matter}
\end{center}

\twocolumngrid
\appendix

\section{Energetics-based computation of the Wigner function of the detected mode}
The Wigner Function of a temporal mode $\hat{b}_v$ of the output field can be
computed following the ``virtual cavity'' method proposed in
Ref.~\cite{kiilerich2020quantum}. In the ``virtual cavity'' method, one defines
a composite system $\rho_{sv}(t)$ of the TLA and an auxiliary bosonic mode, \ie, the ``virtual cavity'' mode, with annihilation operator $\hat{a}_v$. Starting
from $\rho_{sv}(0)=\ketbra{g}{g}_s\otimes\ketbra{\text{vac}}{\text{vac}}_v$, one
writes down and solves a Lindblad master equation for $\rho_{sv}(t)$ such that
$\rho_v(\infty)
= \Tr_s[\rho_{sv}(t=\infty)]$ gives us the state of the temporal mode
$\hat{b}_v$.
However, numerically solving such master equations necessarily requires a
truncation of the infinte dimensional ``virtual cavity'' bosonic Hilbert space
to $\text{Span}\{\ket{\text{vac}}, \ket{1}, \dots \ket{n}\}$.
When $\Tr[\hat{a}^\dagger_v \hat{a}_v
\rho_{sv}(t)]$ is large, such as is the case for large $|\alpha|^2$ input pulses, large $n$ is required and
hence numerical computation is slow.

We circumvent the above problem by first splitting the energy in the virtual
cavity as $\Tr[\hat{a}^\dagger_v \hat{a}_v \rho_{sv}(t)] = | \Tr[ \hat{a}_v
\rho_{sv}(t)]|^2 + \Tr[\delta \hat{a}^\dagger_v \delta \hat{a}_v \rho_{sv}(t)]$,
where the first and second terms are respectively the ``coherent energy''
$E_\mathcal{C}(t)$ and ``incoherent energy'' $E_\mathcal{I}(t)$ of the virtual
cavity. Since $\langle \hat{a}_v(t) \rangle$ can be obtained without solving the
master equation for $\rho_{sv}(t)$ (see below), we can define the ``displaced frame'' state
$\tilde{\rho}_{sv}(t) = \hat{D}^\dagger(\lambda(t)) \rho_{sv}(t)
\hat{D}(\lambda(t))$, where $D(\lambda(t)) = \exp(\lambda(t) \hat{a}_v^\dagger -
\lambda^*(t)a_v)$ and $\lambda(t) = \langle \hat{a}_v(t) \rangle$.  By construction, $\tilde{\rho}_v(t) =
\Tr_s[\tilde{\rho}_{sv}(t)]$ evolves in the same way as $\rho_v(t) =
\Tr_s[\rho_{sv}(t)]$ except that $E_\mathcal{C}(t) = 0$ for all $t$, which
allows us to use a much smaller dimensional truncated Hilbert space.

Explicitly, we define $g_v(t)=-v^*(t)/\sqrt{\int_0^t \dd t^\prime
\,|v(t^\prime)|^2}$, $\hat{H}_{sv}(t) =
\sqrt{\gamma}(g_v(t)\hat{a}_v^\dagger\hat{\sigma} - g^*_v(t)\hat{\sigma}^\dagger
\hat{a}_v)/2i $, $\hat{c}_{sv}(t)=\sqrt{\gamma}\hat{\sigma} +
g_v^*(t)\hat{a}_v$, $\alpha(t) = \alpha u(t)$,
$\hat{H}_{d_s}(t) = i (\alpha^*(t) \hat{\sigma} - \alpha(t) \hat{\sigma}^\dagger)$, and $\lambda(t) =  -e^{-\frac{1}{2} A(t)} \int_{0}^{t} \dd t^\prime g_v(t^\prime) e^{\frac{1}{2}A(t^\prime)} (\alpha(t) + \langle \sigma(t) \rangle)$
where $A(t) = \int_{0}^{t} \dd s \, |g_v(s)|^2$ and $\langle \sigma(t) \rangle$
is obtained by solving the optical Bloch equations for the TLA under the driving Hamiltonian $\hat{H}_{d_s}(t)$ defined above. We also define $l(t) = \alpha(t) g_v(t) + \frac{1}{2} \lambda(t) |g_v(t)|^2 + \dot{\lambda}^*(t)$ and $\hat{H}_{d_v}(t) = i(l^*(t) \hat{a}_v - l(t)\hat{a}_v^\dagger)$. Denoting ${\cal D}[\hat{A}]\bullet = \hat{A}\bullet\hat{A}\dg - \half\{\hat{A}\dg \hat{A},\bullet\}$, we solve~\cite{qutip5} the following Lindblad master equation
\begin{equation}
  \dot{\tilde{\rho}}_{sv} = -i[\hat{H}_{sv}(t) +\hat{H}_{d_s}(t) +\hat{H}_{d_v}(t),\tilde{\rho}_{sv}] + \mathcal{D}[\hat{c}_{sv}(t)]\tilde{\rho}_{sv}
\end{equation}
to obtain $\tilde{\rho}_v(\infty)$. Finally, we recover the correct $E_\mathcal{C}$ by doing $\rho_v(\infty) = \hat{D}(\lambda(\infty))  \tilde{\rho}_v(\infty) \hat{D}^\dagger(\lambda(\infty))$, from which we can easily obtain the Wigner Function in the temporal mode $\hat{b}_v$ through
\beq
W(z) = \frac{1}{\pi^2} \int d^2\mu \, e^{\mu^* z - \mu z^*} \Tr\left[e^{\mu\hat{a}_v - \mu^*\hat{a}_v}\rho_v(\infty)\right]\,.
\eeq
Hence, we see that energy fluctuations are useful not only in characterizing the scattering process through $K^\mathcal{I}(t,t^\prime)$, but also in speeding up numerical computations.

\section{Energy of an output temporal mode $\hat{b}_v$}
The energy $E_v$ of a temporal mode $\hat{b}_v$ can be written as $E_v = \omega_v \langle \hat{b}_v^\dagger \hat{b}_v \rangle$, where $\omega_v = \mel{\text{vac}}{\hat{b}_v \hat{H}_{\rm F} \hat{b}^\dagger_v}{\text{vac}}$ is the effective frequency of the mode $\hat{b}_v$.
Since the temporal profile of our input pulse $u(t)$ is real-valued, the incoherent energy kernel $K^\mathcal{I}(t,t^\prime)$ will also be real-valued, which leads to the eigenfunctions $w_i(t)$ being also real-valued. Since the Fourier transform of a real-valued function is symmetric in $\omega$-space, for each $(\alpha,\Delta)$ when we choose the temporal mode filter function $ w_{\rm inc}(t)$, we have $\omega_{v} = \omega_q$. The same thing occurs for the temporal mode $w_{\rm ng}$, as long as we restrict ourselves to real-valued linear combinations. Hence, we can simply measure the energy of both the input pulse and the output temporal mode in units of $\omega_q$, justifying the treatment of $|\alpha|^2$, $\langle \hat{b}_v^\dagger \hat{b}_v \rangle$, and other photon-number related quantities as energies in this manuscript.  More details can be found in the SM~\cite{referenceToSupplementalMaterial} (See S.5).

\section{Relation between $K^\mathcal{I}(t,t^\prime)$ and the incoherent energy}
The total energy $E_v$ of an output mode $\hat{b}_v$ can be written as
\begin{equation}
  \label{eqn: total energy from K}
  \ex{\hat{b}_v\dg \hat{b}_v} = \iint \dd t \dd t^\prime  \,v^*(t)K(t,t^\prime) v(t^\prime)
\end{equation}
where $K(t,t^\prime) := \langle
\hat{b}^\dagger_\text{out}(t) \hat{b}_\text{out}(t^\prime)\rangle$ is the usual~\cite{gardiner2004quantum,kiilerich2020quantum} quantum optical total
energy kernel. Defining the Heisenberg operator $\delta \hat{b}_\text{out}(t) = \hat{b}_\text{out}(t) - \langle \hat{b}_\text{out}(t)\rangle$, we see that $K^\mathcal{I}(t,t^\prime)$  is related to $K(t,t^\prime)$ by $K^\mathcal{I}(t,t^\prime) = K(t,t^\prime) - \langle \hat{b}^\dagger_\text{out}(t) \rangle \langle \hat{b}_\text{out}(t^\prime) \rangle$. Hence, the incoherent energy $E\incoh = \ex{\delta \hat{b}_v\dg\delta \hat{b}_v}$ of an output mode $\hat{b}_v$ can be written as
\begin{equation}
  \ex{\delta \hat{b}_v\dg\delta \hat{b}_v} = \iint \dd t \dd t^\prime  \,v^*(t)K^\mathcal{I}(t,t^\prime) v(t^\prime).
\end{equation}
We note that $\ex{\delta \hat{b}_v\dg\delta \hat{b}_v}$ takes the largest value
when $v(t)$ is the eigenfunction of $K^\mathcal{I}(t,t^\prime)$ with the largest
eigenvalue. Furthermore, because $K^\mathcal{I}(t, t^\prime) =
\ex{\s{}^\dagger(t) \s{}(t^\prime)} - \ex{\s{}\dg(t)} \ex{\s{}(t^\prime)}$, the
largest eigenvalue $\delta n_0$ of $K^\mathcal{I}(t, t^\prime)$ is less than
$1$, which is a manifestation of the fact that the TLA can scatter at most $1$
photon on average into a single temporal mode. This is why the energetics-based
computation method described above allows for a much smaller dimensional
truncated Hilbert space as compared to the usual ``virtual cavity'' method,
leading to tremendous speed-ups in the numerical computations. Furthermore, $K^\mathcal{I}(t,t^\prime)$ can be experimentally computed with standard resonance fluorescence methods.

\section{Results using the most populated mode}
\begin{figure}[h]
  \centering
  \includegraphics[width=0.49\textwidth, trim={0cm 0cm 0cm 0cm},clip]{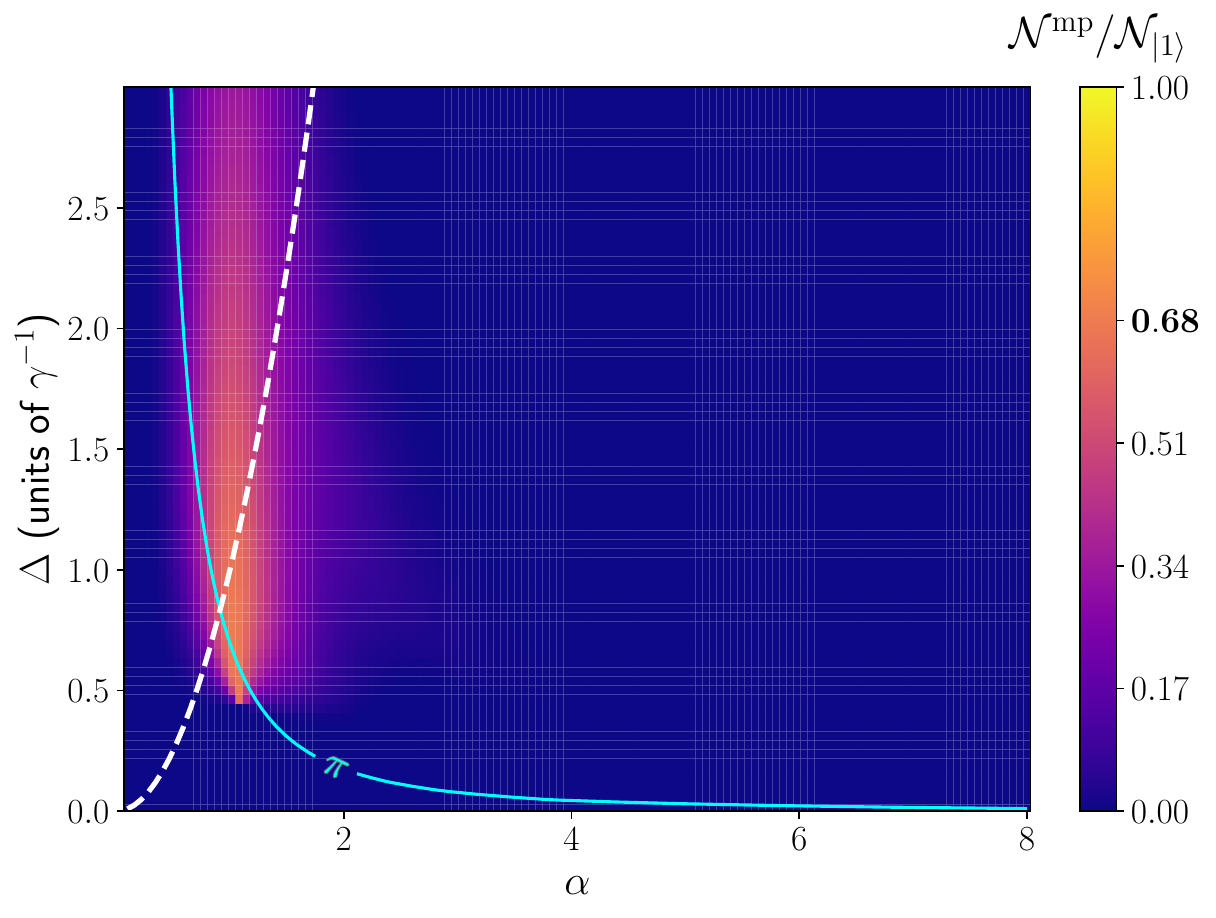}
  \caption[]{$\mathcal{N}^\text{mp}$ of the field in the most
    populated mode $w_\text{mp}$.  The white and cyan contours are as in
  \figref{fig: fig2}. A maximum of $0.68$ is attained for $\mathcal{N}^\text{mp}$ over the $(\alpha,\Delta)$ values considered.}
  \label{fig: wmpResults}
\end{figure}

Denote $w_\text{mp}(t)$ as the eigenfunction of $K(t,t^\prime)$ with the largest eigenvalue and $\mathcal{N}^{\mathrm{mp}}$ as the corresponding WN in that mode. From \eqref{eqn: total energy from K}, we see that choosing $w_\text{mp}$ as the output temporal mode thus corresponds to indiscriminately capturing as much energy as posssible in a single temporal mode. Comparing \figref{fig: wmpResults} with \figref{fig: fig2}(a), we see that using $w_\text{mp}$ causes one to miss out on capturing the single photon that is emitted in the sharp $\pi$-pulse regime (large $\alpha$, small $\Delta$), since in this regime, most of the photons that are emitted by the TLA are coherent photons that would be captured by $w_\text{mp}$. This further motivates the use of $w_\text{inc}$ and subsequently $w_\text{ng}$ for separating the relevant energetic contributions that give rise to WN.

\section{Fine structure of the incoherent energy}
We split the incoherent energy in mode $v$ in the following way:
\begin{equation}
  \label{eqn: energy decomp part1}
  E^v\incoh = E^v\sq +  E^v\mx + \tilde{E}^v\ng \,.
\end{equation}
$E^v\sq = \left(\sqrt{\langle\delta{X_\phi^2}\rangle} - \sqrt{\langle\delta{Y_{\phi}^2}\rangle}\right)^2$, with $X_\phi := (\hat{b}_v e^{-i\phi} + \hat{b}^\dagger_v e^{i\phi})/\sqrt{2}$, $Y_\phi := X_{\phi+\pi/2}$, where $\phi$ is chosen such that $\langle\delta{X_\phi^2}\rangle$ is maximal. $E^v\sq$ captures an energetic contribution due squeezing. $E\mx = n_{th}$, with $n_{th}$ the number of thermal photons in a thermal state that has the same von Neumann entropy as the state of the mode. $E\mx$
captures an energetic contribution due to the mixedness of the field. The remnant contribution to $E\incoh$,
\begin{equation}
  \tilde{E}^v\ng = \sqrt{\left(\ex{\delta \hat{b}_v\dg \delta \hat{b}_v} + \half\right)^2 - |\ex{\delta \hat{b}_v^2}|^2}-\left(n_{th}+\frac{1}{2}\right)\,.
\end{equation}
is an energetic contribution of non-Gaussian effects, like WN.

\section{Other supplementary plots}
\begin{figure}[h]
  \centering
  \includegraphics[width=0.49\textwidth, trim={0cm 0cm 0cm 0cm},clip]{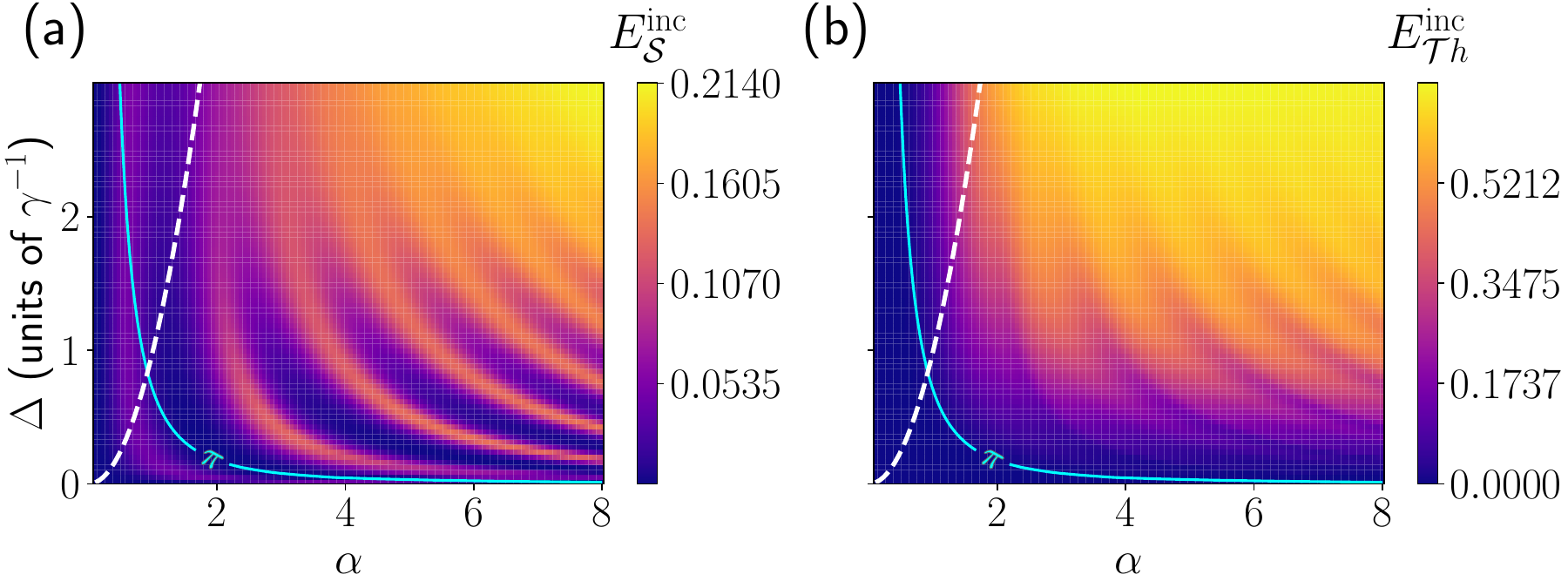}
  \caption[]{ $E^\text{inc}\sq$ and $E^\text{inc}\mx$ of the field in the most
    incoherent mode $w_\text{inc}$.  The white and cyan contours are as in
  \figref{fig: fig2}.  }
  \label{fig: wincEsqAndnth}
\end{figure}
\begin{figure}[h]
  \centering
  \includegraphics[width=0.49\textwidth, trim={0cm 0cm 0cm 0cm},clip]{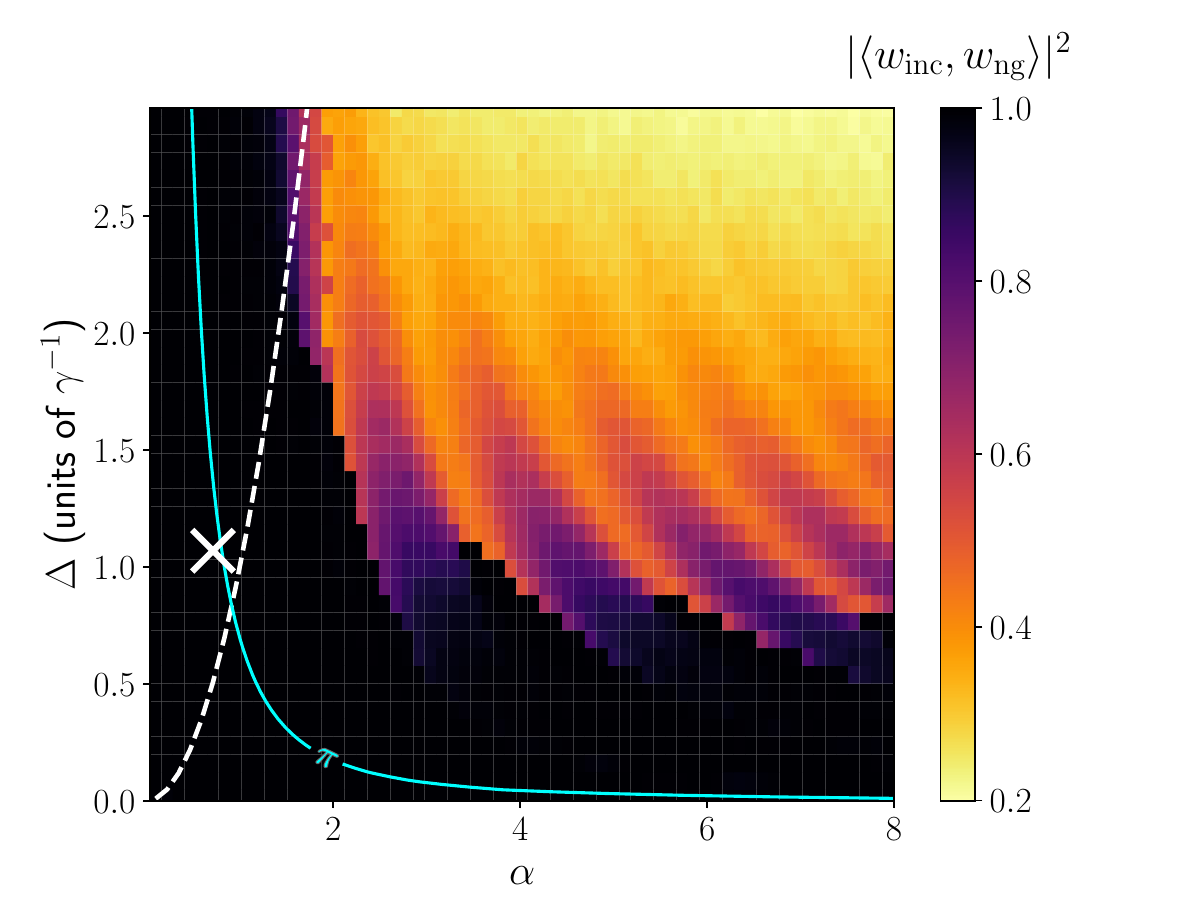}
  \caption[]{$|\langle w_\text{inc}, w_\text{ng} \rangle|^2 := \langle|
    \int_{0}^{\infty} \dd t \, w^*_\text{ng}(t) w_\text{inc}(t)\rangle|^2$ values
    for each $(\alpha, \Delta)$ in \figref{fig: fig3}.  The white and cyan
  contours are as in \figref{fig: fig2}, and the white cross is as in \figref{fig: fig4}.  }
  \label{fig: winc_wng_overlap_squared}
\end{figure}

\figref{fig: wincEsqAndnth} displays $E^\text{inc}\sq$,
$E^\text{inc}\mx$ for the most incoherent mode $w_\text{inc}$. Comparing with \figref{fig: fig2},  as expected, in
high rank regions [${\mathfrak R}_1 \sim 1$], $E^\text{inc}\mx$ and
$E^\text{inc}\sq$ vanish, such that $E^\text{inc}\incoh \sim {\cal
W}^\text{inc}\ng$. Conversely for ${\mathfrak R}_1 <1$, incoherent energy mostly
consists in $E^\text{inc}\mx$, and to a lesser degree $E^\text{inc}\sq$.
\figref{fig: winc_wng_overlap_squared} shows that the $\Delta \tilde{E}\ng
\approx 0$ regions in \figref{fig: fig3}(a) occurs when the the optimization
procedure for $w_\text{ng}(t)$ returns $w_\text{ng} \sim w_\text{inc}$.
Furthermore, we see that the point with the highest $\mathscr{E}$ has
$w_\text{ng} \sim w_\text{inc}$.

\section{Data availability}
The code used to generate these plots can be found in \cite{referenceToCode}.

\clearpage
\onecolumngrid

\begin{center}
  \textbf{\large Supplementary Material: Optimizing Wigner Negativity in Scattering Processes Using Energetic Cost Functions}
\end{center}

\renewcommand{\thefigure}{S\arabic{figure}}
\setcounter{figure}{0}

\renewcommand{\theequation}{S.\arabic{equation}}
\setcounter{equation}{0}

\renewcommand{\thesection}{S.\arabic{section}}
\setcounter{section}{0}

\section{Obtaining a displaced single photon state in the output temporal mode}
\begin{figure}[h]
  \centering
  \includegraphics[width=0.8\textwidth, trim={0cm 0cm 0cm 0cm},clip]{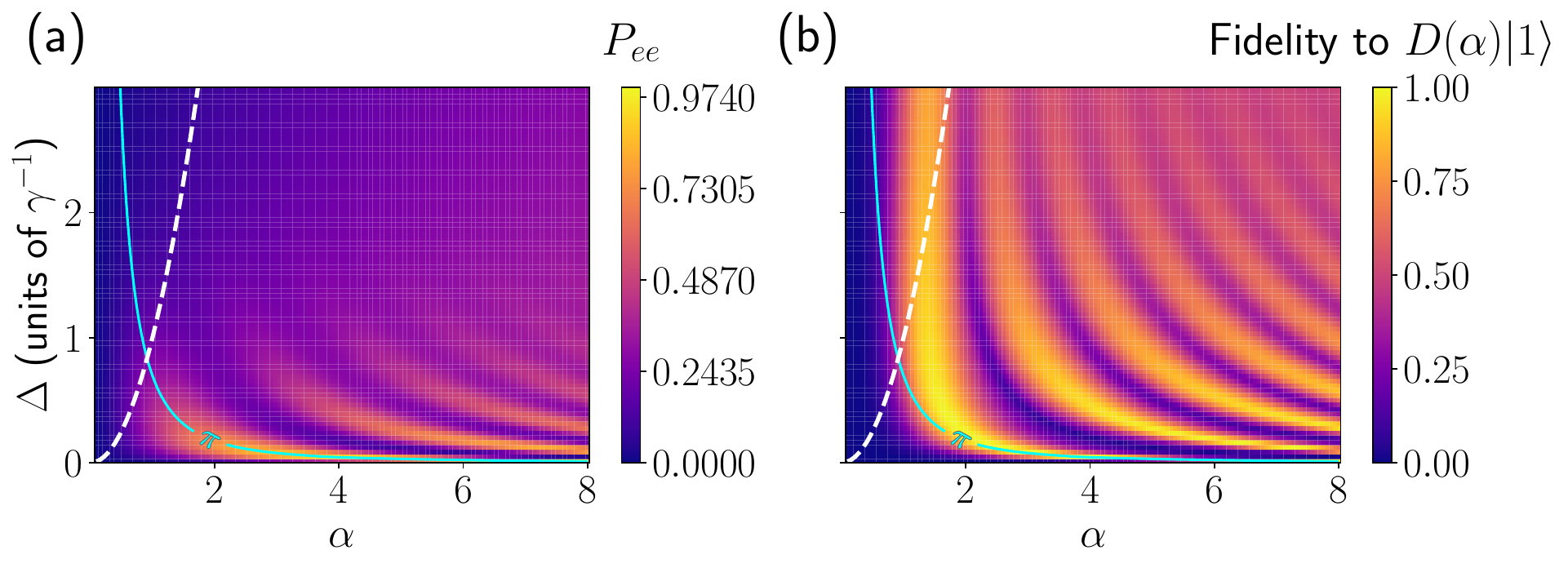}
  \caption[]{\textbf{(a)} $P_{ee}$ of the TLA right after the input pulse. Values closer to $1$ indicate a sharper and hence closer-to-ideal $\pi$-pulse. \textbf{(b)} Fidelity of the field state in the temporal mode $w_\text{inc}$ to that of a displaced single photon state. In both figures, the cyan contour lines represent
    $(\alpha, \Delta)$ values such that the input coherent state executes a finite
  duration $\pi$-pulse. The white contour line separates the regions of $|\alpha|^2 / \gamma \ll \Delta$ (spontaneous emission regime) from the regions where $|\alpha|^2 / \gamma \gg \Delta$ (stimulated emission regime). }
  \label{fig: fidelityToDisplacedSP}
\end{figure}

In Fig.~2 and Fig.~3 of the main text, we find large values of WN when the input pulse approximates a finite-width, finite-intensity $(2k+1)\pi$-pulse, where $k \in {\mathbb Z^+}$. The highest negativities occur for temporally sharper pulses, that is, when $\Delta\ll\gamma$ and $\alpha\gg 1$. In this regime, $w_\text{inc} \sim w_\text{ng}$, and hence we can choose the output temporal mode to be $w_\text{inc}$. Defining $P_{ee} := \langle \sigma^\dagger \sigma \rangle$ as the excited state population of the TLA right after the input pulse, we can see in Fig.~\ref{fig: fidelityToDisplacedSP}(a) that, in this regime, the TLA approximately reaches population inversion $P_{ee} \approx 1$. Hence, the TLA emits a displaced single photon $(D(\alpha)\ket{1})$ with high fidelity in the temporal mode $w_\text{inc}$,  yielding negativities $\mathcal{N}^{\mathrm{inc}} \approx \mathcal{N}_{\ket{1}}$.  While this result is unsurprising, the large input $|\alpha|^2$ is energetically costly.

\begin{figure}[H]
  \centering
  \includegraphics[width=1.0\textwidth, trim={0cm 0cm 0cm 0cm},clip]{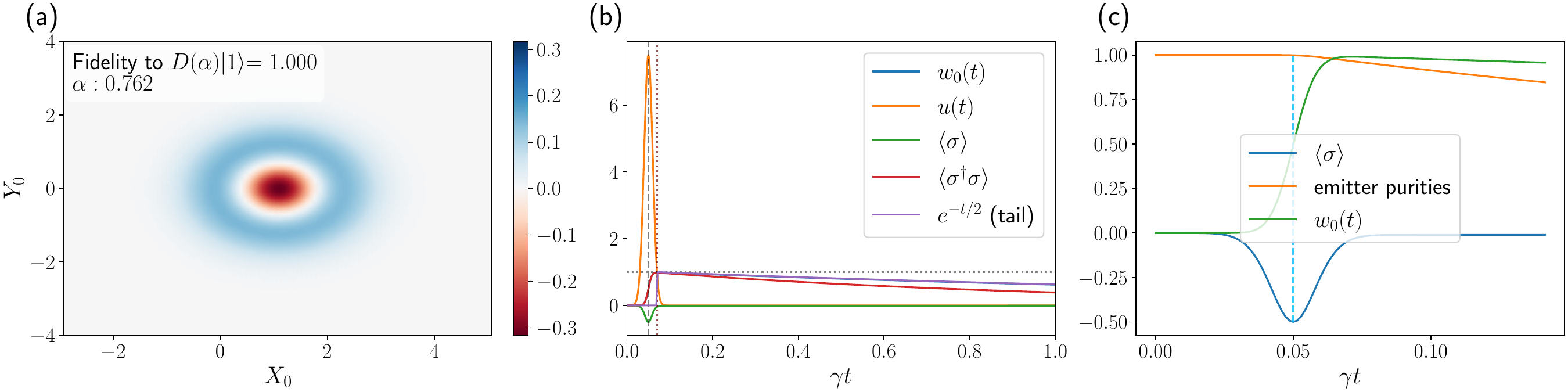}
  \caption[]{Here, we simulate an incident pulse $u(t)$ with  $\alpha = 8.34$
    and $\Delta = 0.01\gamma^{-1}$, which approximates a $\pi$-pulse, on the
    TLA.  \textbf{(a)} The Wigner Function for the field state in the temporal
  mode $\hat{b}_{w_0}$ which has the highest incoherent energy. \textbf{(b)} Various atomic expectation values plotted together with $w_0(t)$ and $u(t)$. As mentioned above, $\langle \sigma \rangle(t) \approx 0$ outside of the input pulse. Furthermore, $w_0(t)$ follows $\langle \sigma^\dagger \sigma \rangle(t)$ in the pulse and it follows $e^{-\gamma t/2}$ outside of the pulse. Here, the exponential tail starts as soon as the emitter is approximately in $\ket{e}$. \textbf{(c)} Similar to \textbf{(b)} but with $\gamma t$ values focused on the duration of the input pulse. We see that the purity of the TLA decreases even during the input pulse, which is because of the finite duration of our pulse.}
  \label{fig: pi pulse analysis}
\end{figure}

The displacement in the single photon state can be easily understood by first computing
\begin{align}
  \langle \hat{b}_{w_0} \rangle
  & = \int_0^{\infty} \dd t \, w_0(t) \langle \hat{b}_\text{out}(t) \rangle                                                                                                                       \\
  & = \int_0^{\infty}  \dd t \, w_0(t) \left(\langle \hat{b}_\text{in}(t) \rangle + \sqrt{\gamma}\langle \sigma(t) \rangle\right)                                                                 \\
  & = \int_{0}^{\infty}  \dd t \, w_0(t) \frac{(2k+1)\sqrt{\pi}}{2\sqrt{2}\Delta} e^{-\frac{(t-t_0)^2}{2\Delta^2}} + \sqrt{\gamma}\int_{0}^{\infty}  \dd t \, w_0(t) \langle \sigma(t) \rangle\,,
\end{align}
where in the last line, we used the fact that the input state performs a $(2k+1)\pi$ pulse on the TLA, {\em i.e.}, that $\int \dd t \alpha u(t) = (2k+1)\frac{\pi}{2}$ for $k \in {\mathbb Z^+}$. In the limit of $\Delta\ll\gamma$ and $\alpha\gg 1$ the Gaussian envelope approximates a Dirac $\delta$-function, yielding $\langle \hat{b}_{w_0} \rangle \approx \frac{(2k+1)\pi}{2} w(t_0)$. This is approximates an ideal $(2k+1)\pi$-pulse. The output temporal mode for the single photon scattered from such a pulse is well-known to be $w_0(t) = \Theta(t - t_0)e^{-\gamma t/2}$, where $\Theta(t - t_0)$ is the Heaviside step function with $\Theta(t_0) = 0$.
An illustration of
the above for a $\pi$-pulse with $\Delta = 0.01$ is shown in Figure~\ref{fig: pi
pulse analysis}. Since we working with a continuous approximation to a $\pi$-pulse, we use continuous functions as output temporal mode functions,which means replacing the Heaviside step function with a straight line interpolation such that $w(t_0) = 1/2$. Hence, we can expect a displacement of $\pi/4\approx 0.785$; a smaller displacement of $0.762$ observed in Figure~\ref{fig: pi pulse analysis} for $\Delta = 0.01$ is due to the finiteness of the pulse. We emphasize that this displacement is merely an artifact of using a straight-line interpolation for the Heaviside step function; in the exact $\pi$-pulse limit, we can use a true Heaviside step function and hence the displacement vanishes.

\section{Details of numerical optimization of \texorpdfstring{$\tilde{E}^{\mathrm{ng}}_{\mathcal{N}\mathcal{G}}$}{Eₙ₉}}
In the main text, it was mentioned that for each $(\alpha, \Delta)$ of the input pulse, we used the functions $w_i(t)$ which are the eigenvalues of $K^\mathcal{I}$ to construct a new mode
\begin{equation}
  \label{eqn: v(t) as a linear combi of Kdelta eigenmodes}
  w_{\alpha, \Delta}(t) = \sum_{j}^{} c_j w_j(t)
\end{equation}
where $c_i$ are chosen such that $w_{\alpha, \Delta}(t)$ is square normalised (in the main text we used the notation $w(t)$; here we add the subscripts to explicitly show the dependence on $(\alpha, \Delta)$). The intuition for this form of $w_{\alpha,\Delta}(t)$ is that only temporal modes with non-zero incoherent energy can have any non-Gaussian energetic contribution, and thus the mode with the largest non-Gaussian energetic contribution should be a linear combination of the modes ${w_i(t)}$ that have non-zero incoherent energy.
Thereafter, we restrict ourselves to real-valued $c_i \in [-2.5, 2.5]$ and we
used a genetic algorithm~\cite{storn1997differential} to find the $c_i$ and
hence the $w_{\alpha,\Delta}(t)$ such that the output field in
$\hat{b}_{w_{\alpha,\Delta}}$ has the highest value of
$\tilde{E}^{\mathrm{ng}}_{\mathcal{N}\mathcal{G}}$.  For each $(\alpha, \Delta)$, we
considered all the eigenfunctions $w_i(t)$ in \eqref{eqn: v(t) as a linear combi
of Kdelta eigenmodes} that have at least $0.01$ incoherent photons. This
neglecting of temporal modes with less than $0.01$ incoherent photons (which is
the finite rank approximation mentioned above), together with the fact that the
optimization does not have a unique maxima, means that the numerical results we
obtained in the main text lower bounds the possible
$\tilde{E}^{\mathrm{ng}}_{\mathcal{N}\mathcal{G}}$ and also the possible $\mathcal{N}^{\mathrm{ng}}$
that can be obtained.

\section{Finite rank approximation of $K^\mathcal{I}(t,t^\prime)$}

\label{appendixSec:finiteRankApprox}
For numerical simulations, it is helpful
to perform a finite rank approximation of $K^\mathcal{I}(t,t^\prime)$ by setting
to zero the eigenvalues which are less than some $\epsilon$ that we can freely
choose. This is mathematically justified since $K^\mathcal{I}(t,t^\prime)$ is compact~\cite{conway2007operators}. This gives us
\begin{align}
  K^\mathcal{I}(t,t^\prime) & = \sum_j \, \delta n_j w_j(t) w^*_j(t^\prime)                                                                  \\
  \label{eqn: finite rank approx}
  & \approx \sum_{j=0}^{m-1}\delta n_j w_j(t) w_j^*(t^\prime) + 0\times w_\perp(t)w_\perp^*(t^\prime),
\end{align}
where in the third line we define $w_\perp(t) = \sum_{j\geq m} \frac{\langle \hat{b}_{w_j}^\dagger \rangle}{\sqrt{\sum_{l \geq m} |\langle \hat{b}_{l} \rangle|^2}} w_j(t)$.
Under these assumptions, we will show, momentarily, that $\langle \hat{b}^\dagger_{w_\perp} \hat{b}_{w_\perp}
\rangle = \sum_{j \geq m} |\langle \hat{b}_{w_j}\rangle|^2$, meaning that the output field in the temporal mode $\hat{b}_{w_\perp}$ is a
coherent state that captures all the coherent energy not in the first $m$ modes.
\eqref{eqn: finite rank approx} can be made
as precise as we wish by choosing $\epsilon$ such that the total number of
output photons is approximately equal to the total number of input photons up to some tolerance $f(\epsilon)$,
that is, as long as $|\langle \hat{b}_{w_\perp} \rangle|^2 +
\sum_{j=0}^{m-1}|\langle \hat{b}_{w_j}\rangle|^2 + \langle \delta \hat{b}^\dagger_{w_j}
\delta \hat{b}_{w_j} \rangle  = |\alpha|^2 - f(\epsilon) \approx |\alpha|^2$.

We now show that the orthogonal mode, under the finite rank approximation, captures a coherent field. Let us define the temporal mode $\hat{b}_{z} = \int \dd t z(t) \hat{b}_{\text{out}}(t)$, where $z(t) = \ex{\hat{b}_{\rm out}\dg(t)}/|\beta|$, and $|\beta|^2 := \int \dd t\, |\ex{\hat{b}_{\rm out}\dg(t)}|^2$.
Applying a displacement of $|\beta|$ on mode $\hat{b}_{z}$, we find $D_{z}(|{\beta}|) \hat{b}_z D^{\dagger}_{z}(|{\beta}|) = \hat{b}_z - |\beta|=\hat{b}_z - \langle b_{z}\rangle$, where $D_{z}(\lambda) = e^{\lambda \hat{b}_z^\dagger - \lambda^* \hat{b}_z}$. By defining the displaced field state $\ket{\text{Field}_D} = D^\dagger_z(|\beta|)\ket{\text{Field}}$, we see that the output energy kernel for this state is equivalent to the incoherent energy kernel, {\em i.e.,}
\begin{align}
  \mel{\text{Field}_D}{\hat{b}_\text{out}^\dagger(t) \hat{b}_\text{out}(t^\prime)}{\text{Field}_D}
  & = \mel{\text{Field}}{\delta \hat{b}_\text{out}^\dagger(t) \delta \hat{b}_\text{out}(t^\prime)}{\text{Field}} \\
  & = K^\mathcal{I}(t, t^\prime) = \sum_{j=0}^{m-1} \delta n_j w_j(t) w_j^*(t^\prime)\,.
\end{align}
Since we have assumed that all modes with $j\geq m$ have zero incoherent energy, this means that these modes of the displaced field state have zero total energy as well. As such, they must be in the vacuum state, {\em i.e.}, $\ket{{\rm Field}_D} = \ket{\Psi}_{w_0,w_1, \dots, w_{m-1}} \otimes \ket{\rm vac}_{w_m, w_{m+1}, \dots} \equiv \ket{\Psi}_{w_0,w_1, \dots, w_{m-1}} \otimes \ket{\rm vac}_{w_\perp}$.

Finally, by defining $\langle f, g\rangle = \int dt \, f^*(t) g(t)$, we can decompose the $z(t)$ mode into the natural orthogonal basis $\{w_j(t)\}_j$ as $z(t) = \sum_{j=0}^{m-1}\langle w_j, z \rangle w_j(t) + \langle w_\perp, z \rangle w_\perp(t)$,
which allows us to divide the displacement operator as
\begin{equation}
  D_z(\lambda) = \prod_{j=0}^{m-1} D_{w_j}(\lambda \langle \beta, w_j \rangle)  D_{w_\perp}(\lambda \langle \beta, w_\perp \rangle)\,.
\end{equation}
Thus, undoing the displacement on the field yields,
\begin{align}
  \ket{\text{Field}}
  & = D_z(|\beta|) \ket{\text{Field}_D}                                                                                   \\
  & = \left(\prod_{j=0}^{m-1} D_{w_j}(\langle \hat{b}_{w_j} \rangle) \ket{\Psi}_{w_1, w_2, \dots w_{m-1}} \right) \otimes
  \left(D_{w_\perp}(\langle \hat{b}_{w_\perp} \rangle)\ket{0}\right)\,,
\end{align}
where $D_{w_\perp}(\langle \hat{b}_{w_\perp} \rangle)\ket{\text{vac}}$ is a coherent state with $|\langle \hat{b}_{w_\perp}\rangle|^2$ number of photons.

\begin{figure}[H]
  \centering
  \includegraphics[width=0.5\textwidth, trim={0cm 0cm 0cm 0cm},clip]{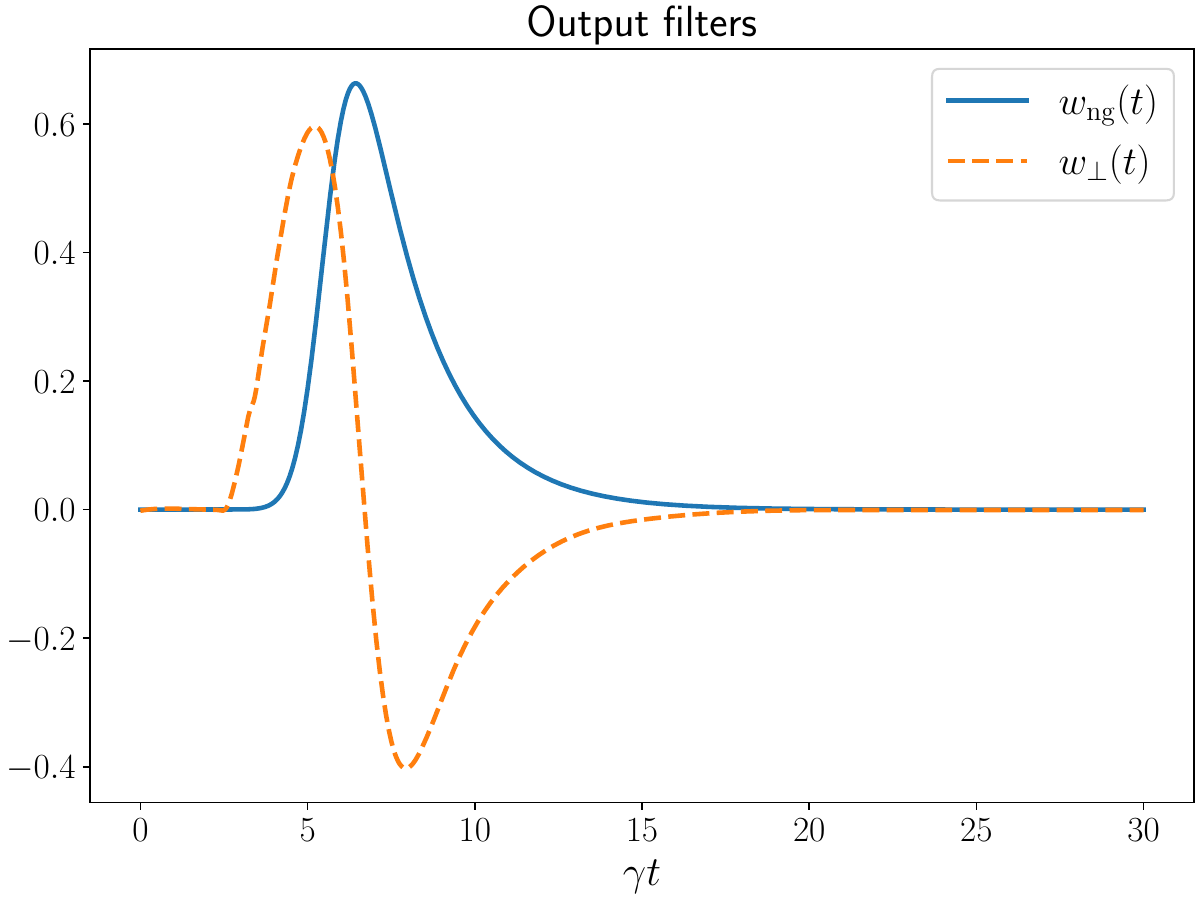}
  \caption[]{
    The temporal modes $w_\text{ng}$ and $w_\perp$.
  }
  \label{fig: most efficient point temporal modes}
\end{figure}

In the main text, we considered the case where $K^\mathcal{I}(t, t^\prime)
\approx \delta n_0 w_0(t)w_0^*(t^\prime) + 0\times
w_\perp(t)w^*_\perp(t^\prime)$, where we can construct $w_\perp(t)$ using the
formula above or equivalently by simply doing $w_\perp(t) \propto \langle
\hat{b}_{\text{out}}\dg(t)\rangle - \langle \hat{b}\dg_{w_0}\rangle w_0(t)$,
where the normalization constant can be found by enforcing $\int \dd t \,
|w_\perp(t)| = 1$. Since there is approximately only one mode with incoherent
energy (and hence non-Gaussian energy), we have $w_\text{ng} = w_\text{inc} =
w_0$. Following the above calculations, this means that
\begin{equation}
  \label{eqn: full output field}
  \ket{\text{Field}} \approx |\psi \rangle_{w_\text{ng}} \otimes |\alpha^\prime \rangle_{w_\perp}
\end{equation}
where $|\alpha^\prime \rangle_{w_\perp}$ is a coherent state with $|\langle
b_{w_\perp}\rangle|^2$ number of photons. The mode functions $w_\text{ng}$ and $w_\perp$ are shown in Figure~\ref{fig: most efficient point temporal modes}.

While the
approximation in \eqref{eqn: full output field} is good,
we can recognize its
limits by noting that the sum of photons in \eqref{eqn: full output field} do
not equal the total number of input photons $|\alpha|^2$, and also that the
field in $\hat{b}_{w_0}$ is only approximately pure ($\ket{\psi}_{w_0}$ has purity $0.977$). This is a result of the finite rank approximation, where we have set $\delta n_i<0.05$ to be zero.

\section{Extra plots at the energetically efficient point}

\begin{figure}[h]
  \centering
  \includegraphics[width=0.8\textwidth, trim={0cm 0cm 0cm 0cm},clip]{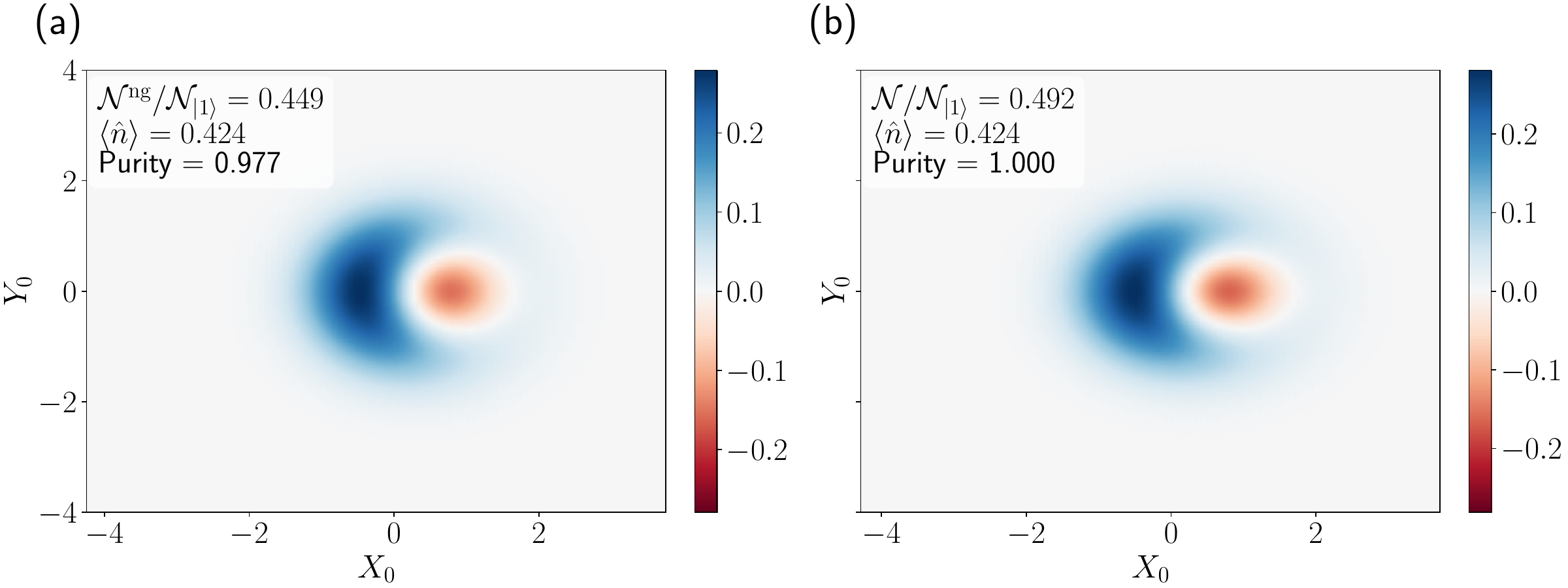}
  \caption[]{\textbf{(a)} The Wigner Function of the field state in the output temporal mode $w_\text{ng} = w_0$ at the energetically efficient point (see main text). \textbf{(b)} The Wigner Function of $\ket{{\rm SNAP}}$ with $\beta$ chosen such $|\beta|^2 = \langle \hat{b}^\dagger_{w_0} \hat{b}_{w_0}\rangle$. The fidelity between the states in the two panels is $0.986$.}
  \label{fig: snapOnVac and energetically efficient WF}
\end{figure}
\begin{figure}[H]
  \centering
  \includegraphics[width=0.4\textwidth, trim={0cm 0cm 0cm 0cm},clip]{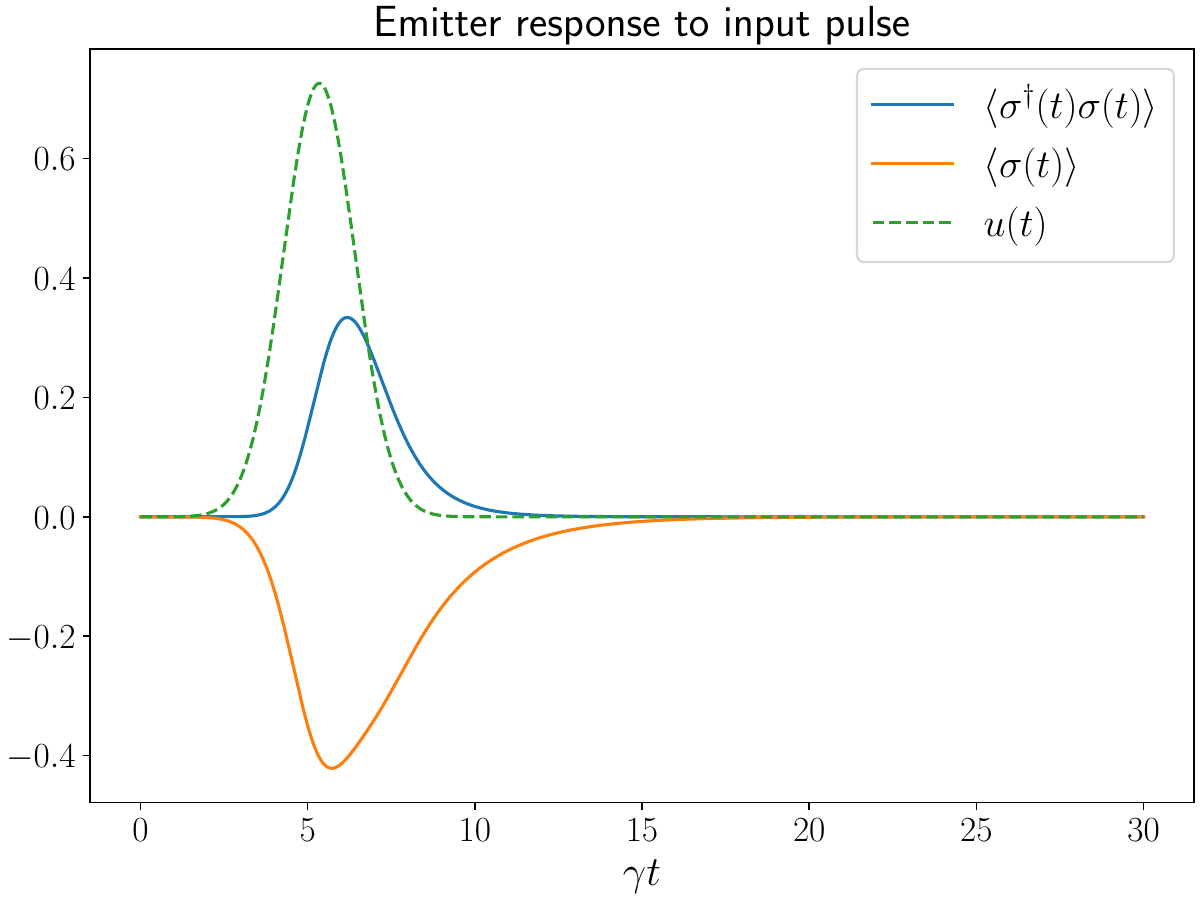}
  \caption[]{The dynamics of the TLA.}
  \label{fig: emitter response}
\end{figure}

First, we see in Fig.~\ref{fig: snapOnVac and energetically efficient WF} that the Wigner functions of both the output field state from the TLA and the state obtained by applying a SNAP gate~\cite{heeres2015snap} on a single mode coherent state, $\ket{{\rm SNAP}} = e^{i\pi \ketbra{0}{0}}\ket{\beta}$ with $|\beta|^2 = \ex{\hat{b}_{w_0}\dg \hat{b}_{w_0}}$, are qualitatively the same, which explains the high fidelity between the states.
Next, in Fig.~\ref{fig: emitter response} we plot various other quantities of
the TLA at the energetically efficient point. We see that for
the TLA, $\langle \sigma^{\dagger}(t)\sigma(t)\rangle < 1/2$, which means that on average, the
TLA does not reach population inversion throughout the scattering process at the energetically efficient point. This makes sense since the input coherent state at that point has less than one input photon. Furthermore, we see the finite non-zero emitter response $\langle \hat{\sigma}(t) \rangle$ during the duration of the input pulse, which gives rise to a displacement in the output field~\cite{prasad2026thermodynamics}.

\section{Energy of the scattered modes}
In this section we show that the energy of both the input and output modes are directly proportional to their number operators, with the constant of proportionality being the characteristic frequency of the TLA, $\omega_q$.

First, let us begin with the input mode. The energy of the input field is determined via the Hamiltonian $\hat{H}_{\rm F, in} = \int \dd \omega\, \omega \hat{b}^\dagger_i(\omega) \hat{b}_i(\omega)$, where $\hat{b}_{i}(\omega)$ denotes the annihilation operator in the Heisenberg picture before the input light pulse has
interacted with the TLA. The input field annihilation operator, as standard in input-output theory \cite{gardiner1985input}, is defined as $\hat{b}_\text{in}(t) = \int \frac{dt}{\sqrt{2\pi}} e^{-i(\omega - \omega_q)(t-t_i)}\hat{b}_i(\omega)$. Also, recall that the mode of the input pulse is defined via $\hat{b}_u = \int \dd t\, u(t) \hat{b}(t)$. Note, here it is not necessary to assume the Gaussian form of $u(t)$ we used in the main text, instead, any real-valued $u(t)$ will suffice. The energy of the input state $\ket{\alpha_{\text{in}}} = e^{\alpha^* \hat{b}_u  - \alpha \hat{b}_u^\dagger} \ket{\text{vac}}$ is
\begin{align}
  \mel{\alpha_\text{in}}{H_{\rm F, in}}{\alpha_\text{in}}
  & = \int d\omega \, \iint \frac{\dd t \dd t^\prime}{2\pi} \omega e^{-i(\omega - \omega_q)(t-t^\prime)}  \mel{\alpha_\text{in}}{\hat{b}^\dagger_\text{in}(t)\hat{b}_\text{in}(t^\prime)}{\alpha_\text{in}} \\
  & = |\alpha|^2 \int d\omega \, \omega  e^{-i(\omega - \omega_q)(t-t^\prime)} \iint \frac{\dd t \dd t^\prime}{2\pi} u^*(t) u(t^\prime)                                                          \\
  & = |\alpha|^2\int d\omega \, \omega |\tilde{u}(\omega - \omega_q)|^2                                                                                                                        \\
  & = \omega_q |\alpha|^2                                                                                                                                                                      \\
  & = \omega_q \ex{\hat{b}_u\dg \hat{b}_u}\,.
\end{align}
where, $\tilde{u}(\omega) = \frac{1}{\sqrt{2\pi}}\int\dd t u(t)e^{-i\omega t}$ and we have used the fact that the Fourier transform of a real-valued function is symmetric.

Now, let us consider the output modes. The energy of the output field is determined via the Hamiltonian $\hat{H}_{\rm F, out} = \int \dd \omega\, \omega \hat{b}^\dagger_f(\omega) \hat{b}_f(\omega)$, where $\hat{b}_{f}(\omega)$ denotes the annihilation operator in the Heisenberg picture after the light pulse has finished interacting with the TLA. The output field annihilation operator is defined as $\hat{b}_\text{out}(t) = \int \frac{dt}{\sqrt{2\pi}} e^{-i(\omega-\omega_q)(t-t_f)}\hat{b}_f(\omega)$. Let $\{w_k(t)\}_k$ be the eigenfunctions of $K^\mathcal{I}(t,t^\prime)$, and consider the temporal mode functions $\{w_k(t)\}_k$ which form an orthonormal basis, {\em i.e.}, $\int \dd t w_k^*(t) w_\ell(t) = \delta_{k\ell}$. Note, $\{w_k(t)\}_k$ are real-valued as the kernel is both real-valued and symmetric. The full output field after the interaction with the TLA can be written as
\begin{equation}
  \ket{\text{Field}} = \sum_{n_1, n_2, \dots} c_{n_1,n_2,\dots} \ket{n_1}_{w_1} \otimes \ket{n_2}_{w_2}\dots\,,
\end{equation}
where $\ket{n}_{w_k} = \frac{(\hat{b}^\dagger_{w_k})^{n}}{\sqrt{n}}\ket{\text{vac}}$ and $\hat{b}_{w_k} = \int \dd t w_k(t) \hat{b}_\text{out}(t)$. The total energy of the output field is
\begin{align}
  \mel{\text{Field}}{\hat{H}_{\rm F, out}}{\text{Field}}
  & = \int \dd \omega\, \omega \ex{\hat{b}^\dagger_f(\omega) \hat{b}_f(\omega)}                                                                                                        \\
  & = \sum_{k,\ell} \int \dd \omega\, \iint \frac{\dd t \dd t^\prime}{2\pi} e^{-i(\omega - \omega_q)(t -t^\prime)} w_k(t)w^*_\ell(t^\prime)\ex{\hat{b}^\dagger_{w_\ell} \hat{b}_{w_k}} \\
  \label{eqn: nice equation for H_F out}
  & = \sum_{k,\ell} \Omega_{k\ell} \ex{\hat{b}^\dagger_{w_k} \hat{b}_{w_\ell}} ,
\end{align}
where $\Omega_{k\ell} = \int \dd \omega \, \omega \tilde{w}_k(\omega - \omega_q) \tilde{w}^*_\ell(\omega - \omega_q)$.

Simplifying $\Omega_{k\ell}$, we obtain
\begin{align}
  \Omega_{k\ell}
  & = \int \dd \omega \, \omega \tilde{w}_k(\omega - \omega_q) \tilde{w}^*_\ell(\omega-\omega_q)                                                                                                                                                                                                     \\
  & = \omega_q\int \dd \nu \, \tilde{w}_k(\nu) \tilde{w}^*_\ell(\nu) +
  \int \dd \nu \, \nu \tilde{w}_k(\nu) \tilde{w}^*_\ell(\nu)                                                                                                                                                                                                                                       \\
  & = \omega_q\delta_{k\ell} + \int \dd \nu \, \nu \left[\underbrace{\frac{\tilde{w}_k(\nu) \tilde{w}_\ell^*(\nu)+\tilde{w}_k^*(\nu) \tilde{w}_\ell(\nu)}{2}}_{\text{Even function in }\nu}+ \frac{\tilde{w}_k(\nu) \tilde{w}_\ell^*(\nu)-\tilde{w}_k^*(\nu) \tilde{w}_\ell(\nu)}{2 }\right] \\
  \label{eqn: Omega_ij vanishing part 1}
  & =  \omega_q\delta_{k\ell} +\int \dd \nu \, \nu \left[ \frac{\tilde{w}_k(\nu) \tilde{w}_\ell^*(\nu)-\tilde{w}_k^*(\nu) \tilde{w}_\ell(\nu)}{2}\right]
\end{align}
where, in the second line, we have defined $\nu = \omega - \omega_q$ and used the fact that $\tilde{w}_k(\omega) = \tilde{w}_k^*(-\omega)$, since $w_k(t)$ are real. Using the above result, we can check that
\begin{align}
  \sum_{k \neq \ell} \Omega_{k\ell} \ex{\hat{b}^\dagger_{w_k} \hat{b}_{w_\ell}}
  & = \sum_{k < \ell} (\Omega_{k\ell} + \Omega_{\ell k}) \ex{\hat{b}^\dagger_{w_k} \hat{b}_{w_\ell}} \nonumber \\
  & = 0.
\end{align}
As such, the energy of the output mode is given by
\begin{equation}
  \mel{\text{Field}}{\hat{H}_{\rm F, out}}{\text{Field}} =\omega_q \sum_k \langle \hat{b}^\dagger_{w_k} \hat{b}_{w_k} \rangle\,.
\end{equation}
Importantly, we can see that each mode $\hat{b}_{w_k}$ contributes to the total energy by their photonic occupation, with each mode having the same proportionality being $\omega_q$. We do consider linear combinations of these mode functions to construct $w_{\rm ng}(t) = e^{i \omega_q t} \sum_k c_k w_k(t)$, the above analysis also holds provided the coefficients $c_k$ are real, which is the case in the main text.

\end{document}